\documentclass[12pt]{article}
\usepackage{amssymb}
\usepackage{graphicx}
\usepackage{mathrsfs}
\font\blackboard=msbm10 at 12pt
\font\blackboards=msbm7
\font\blackboardss=msbm5
\newfam\black
\textfont\black=\blackboard
\scriptfont\black=\blackboards
\scriptscriptfont\black=\blackboardss

\newcommand{\junk}[1]{}

\newcommand{\ba}{\begin{array}}
\newcommand{\ea}{\end{array}}
\newcommand{\be}{\begin{equation}}
\newcommand{\ee}{\end{equation}}
\newcommand{\bea}{\begin{eqnarray}}
\newcommand{\eea}{\end{eqnarray}}
\newcommand{\beas}{\begin{eqnarray*}}
\newcommand{\eeas}{\end{eqnarray*}}

\def\laplace{{\kern1pt\vbox{\hrule height 1.2pt\hbox{\vrule width
1.2pt\hskip
  3pt\vbox{\vskip 6pt}\hskip 3pt\vrule width 0.6pt}\hrule height
  0.6pt}
  \kern1pt}}
\def\scriptlap{{\kern1pt\vbox{\hrule height 0.8pt\hbox{\vrule width
  0.8pt
  \hskip2pt\vbox{\vskip 4pt}\hskip 2pt\vrule width 0.4pt}\hrule height
  0.4pt}
  \kern1pt}}

\def\roughly#1{\raise.3ex\hbox{$#1$\kern-.75em\lower1ex\hbox{$\sim$}}}

\newcommand{\gone}[1]{}

\begin{document}
\pagestyle{plain}
\setcounter{page}{1}

\baselineskip16pt

\begin{titlepage}

\begin{flushright}

\end{flushright}
\vspace{8 mm}

\begin{center}

{\Large \bf Radial quantization of the $3d$ CFT and the higher spin/vector model duality  \\}

\end{center}

\vspace{7 mm}

\begin{center}

{\bf Shan Hu $^{1}$, and Tianjun Li $^{1,2}$}

\vspace{3mm}
{\small \sl $^{1}$State Key Laboratory of Theoretical Physics,
 Institute of Theoretical Physics, } \\
{\small \sl  Chinese Academy of Sciences, Beijing, 100190, P. R. China} \\
{\small \sl $^{2}$School of Physical Electronics,
University of Electronic Science and Technology of China,} \\

{\small \sl Chengdu 610054, P. R. China} \\

\vspace{3mm}

\end{center}

\vspace{8 mm}
\begin{abstract}

We study the radial quantization of the $3d$ $O(N)$ vector model. 
We calculate the higher spin charges whose commutation relations give 
the higher spin algebra. The Fock states of higher spin gravity 
in $AdS_{4}$ are realized as the states in the $3d$ CFT. The inner products 
between these states encode dynamical information. The construction of 
bulk operators from CFT is discussed as well. This serves as the simplest 
explicit demonstration of the CFT definition for the quantum gravity.

\end{abstract}

\vspace{1cm}

\begin{flushleft}

\end{flushleft}
\end{titlepage}
\newpage

\section{Introduction}

Higher spin gauge theory in AdS was conjectured to be dual to the free CFT \cite{01qa,01qb}. 
In \cite{01q}, it was further proved that imposing the higher spin symmetry makes the CFT free. Due to the higher spin symmetry, the two sides of the duality are easy to approach. On CFT side, the free theory is totally solvable and so, the N-point correlation functions for conserved currents can be calculated directly \cite{01qc,01qf}. On gauge theory side, higher spin symmetry also makes it possible to get the exact correlators from the bulk Vasiliev theory \cite{01qc}-\cite{01qe}.

Aside from the identification of the correlation functions, another manifestation of the AdS/CFT duality is the state correspondence \cite{ft67y}. In most AdS/CFT examples, the dual CFT is interacting, for which, it is difficult to construct the states corresponding to the Fock states of the AdS theory. For the higher spin/vector model duality, the CFT is free, making it possible to do a complete radial quantization to get the states and charges which are supposed to be in one-to-one correspondence with the states and charges in the higher spin gravity. In this sense, CFT gives a definition of the quantum higher spin gravity in AdS. A consistent interacting theory for the non-Abelian higher spin gauge theory in AdS has been developed in \cite{1qw}-\cite{6qw}, but the quantization of the theory is still an open problem. With the higher spin gravity quantized, we can then compare the gauge theory results with the CFT predictions.

The higher spin gauge theory is most elegantly formulated in the fame-like formalism with fields taking values in the higher spin algebra \cite{6qw6,6qw7}. However, it is the totally symmetric traceless tensors $h_{i_{1}\cdots i_{s}}$ that have the operator correspondence in CFT. As a result, CFT can only give a definition of the higher spin gauge theory in the metric-like formalism \cite{36q,37q}. Moreover, the CFT realization of the higher spin gravity states automatically obeys some kind of ``Coulomb gauge''. Therefore, to make the comparison between the two sides of the duality, it is also necessary to quantize the higher spin gravity in ``Coulomb gauge''. Since the gauge condition is intrinsically imposed in CFT, the constructed states are all physical without the redundant degrees of freedom to remove. Also, since the local gauge symmetry does not exist anymore, the subtleties of the quantum gravity coming from the diffeomorphism invariance do not arise in its CFT definition. The conclusion also holds for the generic AdS/CFT correspondence without the higher spin symmetry.

In the free $O(N)$ vector model, one can use the higher spin charges and their commutators to get the higher spin algebra. The higher spin charges have a simple oscillator expression $\sum a^{+}a$. We compute two sets of charges $P^{s}_{\mu_{1}\cdots\mu_{n}}$ and $K^{s}_{\mu_{1}\cdots\mu_{n}}$ explicitly, which are the higher spin generalizations of the translation generator $P^{2}_{\mu}=P_{\mu}$, the special conformal transformation generator $K^{2}_{\mu}=K_{\mu}$, and the dilaton $D=P^{2}=K^{2}$. $K^{s}_{\mu_{1}\cdots\mu_{n}} = -(P^{s}_{\mu_{1}\cdots\mu_{n}})^{+}$. $\{P^{s}_{\mu_{1}\cdots\mu_{n}}|s=2,4,\cdots;n=0,1,\cdots,s-1\}$ and $\{K^{s}_{\mu_{1}\cdots\mu_{n}}|s=2,4,\cdots;n=0,1,\cdots,s-1\}$ are two subalgebras of the higher spin algebra.

Single and the multiple particle states are all constructed, composing the complete Fock space for the non-perturbative higher spin gravity in $AdS_{4}$. The 1-particle Hilbert space of the higher spin gravity forms the irreducible representation of the higher spin algebra. The generalized Flato-Fronsdal theorem is re-proved in $AdS_{D+1}$. Since the CFT is free, states with the different numbers of the particles are orthogonal to each other. Single particle states with the different spins are also orthogonal to each other, which is traced to the $SO(3,2)$ symmetry in AdS and CFT. The inner products for the multi-particle states encode the dynamical information. For example, the $\textnormal{spin}\; 0 + \textnormal{spin} \;0 \longleftrightarrow \textnormal{spin} \;0 + \textnormal{spin} \;2$, $\textnormal{spin}\; 4 + \textnormal{spin} \;0 \longleftrightarrow \textnormal{spin} \;2 + \textnormal{spin} \;2$ spin changing processes can happen with the $1/N$ suppression, indicating that the dual AdS theory is not free with the coupling constant $g \sim 1/\sqrt{N}$.

With the CFT given, it is natural to ask aside from the charges, the spectrum and the corresponding states, whether there is more information about the AdS, like the bulk operators, that can be extracted from CFT. The topic has been intensively discussed in the literatures, see, for example, \cite{1qwa}-\cite{qwert1}. We will add more comments. For the AdS theory, let $\Phi(x,\rho)$ be a set of operators in AdS. $\Phi(x,\pi/2)$ are operators in $\partial AdS$. The $SO(3,2)$ transformation law for $\Phi(x,\pi/2)$ is the same as the conformal transformation law of the primary operators $O(x)$ in CFT with the definite conformal dimension. So with the conformal generators in CFT identified with the $so(3,2)$ generators in AdS, $O(x)$ could be taken as the boundary limit of some AdS operators. Usually, with a bulk operator $\Phi(o)$ given, $\{g\Phi(o)g^{-1}|g \in SO(3,2)\}$ could give the whole set of operators in AdS. However, this does not hold for the boundary operators. $\{g\Phi(b)g^{-1}|g \in SO(3,2)\}$ for $b \in \partial AdS$ only gives operators in the boundary. As a result, two set of AdS operators may have the same boundary value, i.e. 
\begin{equation}
	\{\Phi(x,\rho)|(x,\rho) \in AdS_{4}\} \neq \{\Phi'(x,\rho)|(x,\rho) \in AdS_{4}\},
\end{equation}
but
\begin{equation}
	\{\Phi(x,\rho)|(x,\rho) \in \partial AdS_{4}\} = \{\Phi'(x,\rho)|(x,\rho) \in \partial AdS_{4}\}.
\end{equation}
Especially, with the boundary $\Phi(x,\pi/2)$ given, one can always find a consistent bulk extension $\Phi^{f}(x,\rho)$ by solving the free field equation with $\Phi(x,\pi/2)$ the boundary value. So, the bulk extension of $O(x)$ is not unique. We are interested in the Heisenberg operators $\Psi(x,\rho)$ for fields in AdS theory. Even if $O(x)$ could be taken as the boundary limit of $\Psi(x,\rho)$, the CFT data is not enough to determine the bulk $\Psi(x,\rho)$. We may get $O^{f}(x,\rho)$ by solving the free field equation, but obviously, $O^{f}(x,\rho) \neq \Psi(x,\rho)$ unless the theory is free. In principle, once a single bulk $\Psi(o)$ gets the CFT realization, the action of $SO(3,2)$ will then offer the whole set of $\Psi(x,\rho)$ in CFT. However, it is not quite clear which operator in CFT should be identified with $\Psi(o)$. The requirement of $[K_{\mu}-iP_{\mu},\Psi(o)]=0$ and $-i[M_{\mu\nu},\Psi(o)]=\Sigma_{\mu\nu}\Psi(o)$ could guarantee that $\Psi(o)$ is an operator in the bulk with the spin $s$, but is far from enough to uniquely fix it.

Nevertheless, in AdS theory, aside from $\Psi(x,\rho)$, the other interesting operators are $a^{+}_{\omega,j;\lambda,s}$ and $\Phi^{+}_{\lambda,s}(x,\rho)$, which are the operators creating the physical particles of the type $(\lambda,s)$ in momentum space and AdS. Let $a^{+}_{\lambda,s;\lambda,s}$ be the lowest energy operator with $[K_{\mu},a^{+}_{\lambda,s;\lambda,s}]=0$ and $[H,a^{+}_{\lambda,s;\lambda,s}]=-i \lambda a^{+}_{\lambda,s;\lambda,s}$, $\{g  a^{+}_{\lambda,s;\lambda,s} g^{-1}|g \in SO(3,2)\}$ gives $\Phi^{+}_{\lambda,s}(x,\pi/2)$. The bulk extension $\Phi^{+}_{\lambda,s}(x,\rho)$ can be obtained by solving the free field equation as is required by the $SO(3,2)$ symmetry. In CFT, we have $O_{\lambda,s}^{+}(0)=a^{+}_{\lambda,s;\lambda,s}$, from which, one may get the CFT realization of $\Phi^{+}_{\lambda,s}(x,\rho)=O_{\lambda,s}^{+}(x,\rho)$. To calculate the transition amplitude for physical particles, $\Phi^{+}_{\lambda,s}(x,\rho)$ is enough. In fact, in a non-perturbative treatment of the theory in AdS, we start from $\Psi(x,\rho)$ for bare fields but still end up with $\Phi^{+}_{\lambda,s}(x,\rho)$, since it is $\Phi^{+}_{\lambda,s}(x,\rho)$ that is directly related with the observation. In this sense, CFT can indeed provide the complete physical dynamical information in AdS.

The rest of the paper is organized as follows. In section 2, we discuss the radial quantization of the $3d$ free $O(N)$ vector model; in section 3, the $so(3,2)$ generators of the CFT are calculated; in section 4, the higher spin charges are computed, while the corresponding higher spin algebra is considered; in section 5, the one-to-one correspondence between the Hilbert space of the $3d$ CFT and the quantum higher spin gravity in $AdS_{4}$ is constructed; in section 6, we discussed the possibility to get the bulk operators from CFT; the conclusion is in section 7.

\section{The radial quantization of the $3d$ $O(N)$ vector model }

The action of the free scalar field theory with the $O(N)$ symmetry in $3d$ is
\begin{equation}
	S = \int d^{3}x \; \partial_{\mu}\phi \partial^{\mu}\phi,
\end{equation}
where $\phi$ are $N$ real scalars in the fundamental representation of $O(N)$. The field equation is 
\begin{equation}
	\partial^{2}\phi = 0.
\end{equation}
With $x_{1}$ replaced by $i x_{1}$, we get the theory in Euclidean space. In spherical coordinate, $(x_{1},x_{2},x_{3})\rightarrow (r,\theta,\varphi)$, where $r^{2} = x^{2}_{1}+x^{2}_{2}+x^{3}_{3}$. Also, $n_{\mu} = x_{\mu}/r$ with
\begin{equation}
	n_{1} = \cos \theta,\;\;\;\; n_{2} = \sin \theta \cos\varphi,\;\;\;\;n_{3} = \sin \theta \sin\varphi.
\end{equation}
Introducing the dimensionless field $\chi = r^{\frac{1}{2}}\phi$,\footnote{In $2d$, $\chi = \phi$; in $4d$, $\chi =r \phi$.} The solution can be expanded as \cite{1q} 
\begin{equation}\label{3we}
\chi (r,\theta,\varphi)=	\sum_{l\geq 0,-l \leq m \leq l}\; [a_{l,m}\frac{r^{-(l+\frac{1}{2})}}{\sqrt{2l+1}}Y^{*}_{l,m}(\theta,\varphi) +  a^{+}_{l,m}\frac{r^{l+\frac{1}{2}}}{\sqrt{2l+1}}Y_{l,m}(\theta,\varphi)   ]~,~
\end{equation}
\begin{equation}
	\chi^{+} (r,\theta,\varphi)=\chi (\frac{1}{r},\theta,\varphi). 
\end{equation}
Let $r = e^{it}$ with $t$ the proper time, (\ref{3we}) becomes \cite{1q}
\begin{equation}
\chi (r,\theta,\varphi)=	\sum_{l\geq 0,-l \leq m \leq l}\; [a_{l,m}\frac{e^{-i(l+\frac{1}{2})t}}{\sqrt{2l+1}}Y^{*}_{l,m}(\theta,\varphi) +  a^{+}_{l,m}\frac{e^{i(l+\frac{1}{2})t}}{\sqrt{2l+1}}Y_{l,m}(\theta,\varphi)           ]~,~
\end{equation}
with $a_{l,m} = [a^{+}_{l,m}]^{+}$.

Note that
\begin{equation}
	Y_{i_{1}i_{2}\cdots i_{l}} = \frac{x_{i_{1}}x_{i_{2}}\cdots x_{i_{l}} - \textnormal{traces}}{r^{l}}, 
\end{equation}
$\chi$ can also be expanded as 
\begin{equation}
\chi (r,\theta,\varphi)=	\sum \; [a_{i_{1}i_{2}\cdots i_{l}} \frac{r^{-(l+\frac{1}{2})}}{\sqrt{2l+1}}Y_{i_{1}i_{2}\cdots i_{l}} +  a^{+}_{i_{1}i_{2}\cdots i_{l}} \frac{r^{l+\frac{1}{2}}}{\sqrt{2l+1}}Y_{i_{1}i_{2}\cdots i_{l}}         ].
\end{equation}   
Correspondingly,
\begin{equation}\label{rf}
\phi (r,\theta,\varphi)=	\sum \; [a_{i_{1}i_{2}\cdots i_{l}} \frac{r^{-(l+1)}}{\sqrt{2l+1}}Y_{i_{1}i_{2}\cdots i_{l}} +  a^{+}_{i_{1}i_{2}\cdots i_{l}} \frac{r^{l}}{\sqrt{2l+1}}Y_{i_{1}i_{2}\cdots i_{l}}          ]. 
\end{equation}
$a_{i_{1}i_{2}\cdots i_{l}}$ and $a^{+}_{i_{1}i_{2}\cdots i_{l}}$ are traceless totally symmetric
operators encoding $2l+1$ degrees of freedom, the same as $a_{lm}$ and $a^{+}_{lm}$.

If
\begin{equation}
Y_{i_{1}i_{2}\cdots i_{l}}(n_{\mu})  = \sum_{m}	\; f_{i_{1}i_{2}\cdots i_{l}}^{l,m} Y_{l,m}(\theta,\varphi),
\end{equation}
there will be 
\begin{equation}
	a^{+}_{l,m} = \sum \; f_{i_{1}i_{2}\cdots i_{l}}^{l,m} a^{+}_{i_{1}i_{2}\cdots i_{l}},
\end{equation}
\begin{equation}
a_{l,m} = \sum \; (f_{i_{1}i_{2}\cdots i_{l}}^{l,m})^{*} a_{i_{1}i_{2}\cdots i_{l}}.	
\end{equation}
If
\begin{equation}
	a^{+}_{i_{1}i_{2}\cdots i_{l}}=\sum_{m}\; g^{l,m}_{i_{1}i_{2}\cdots i_{l}}a^{+}_{l,m},  
\end{equation}
\begin{equation}
	a_{i_{1}i_{2}\cdots i_{l}}=\sum_{m}\; (g^{l,m}_{i_{1}i_{2}\cdots i_{l}})^{*}a_{l,m},  
\end{equation}
we may have 
\begin{equation}\label{o1}
	\sum \; f_{i_{1}i_{2}\cdots i_{l}}^{l,m}  g^{l,m'}_{i_{1}i_{2}\cdots i_{l}}=\delta_{m,m'}. 
\end{equation}
Also, there are
\begin{equation}
	[a^{k}_{l,m},a^{+k'}_{l',m'}]=\delta_{l,l'}\delta_{m,m'}\delta_{k,k'}, 
\end{equation}
\begin{equation}
	[a^{k}_{i_{1}\cdots i_{l}},a^{+k'}_{j_{1}\cdots j_{l}}]=\delta_{k,k'}\sum_{m}\;  (g^{l,m}_{i_{1}i_{2}\cdots i_{l}})^{*} g^{l,m}_{i_{1}i_{2}\cdots i_{l}},
\end{equation}
where $k,k'=1,\cdots,N$ are color indices. Let
\begin{equation}\label{o2}
\int_{S^{2}} d\Omega \; Y_{i_{1}\cdots i_{l}}(n_{\mu}) Y_{j_{1}\cdots j_{l}}(n_{\mu}) = \sum_{m}	\; f_{i_{1}\cdots i_{l}}^{l,m} (f_{j_{1}\cdots j_{l}}^{l,m})^{*} = N_{i_{1}\cdots i_{l};\;j_{1}\cdots j_{l}}. 
\end{equation}
$a^{+}_{j_{1}\cdots j_{l}}$ is traceless, so 
\begin{eqnarray}
\nonumber N_{j_{1} \cdots j_{l};\;i_{1} \cdots i_{l}}a^{+}_{j_{1}\cdots j_{l}} &=&  \int_{S^{2}} d\Omega \; (n_{j_{1}}\cdots n_{j_{l}}n_{i_{1}}\cdots n_{i_{l}})a^{+}_{j_{1}\cdots j_{l}}    \\\nonumber &=& \frac{l!}{(2l+1)!!}a^{+}_{i_{1}\cdots i_{l}}=\sum_{m}\;(f^{l,m}_{i_{1} \cdots i_{l}})^{*}a^{+}_{l,m}\\&=& \sum_{m}\;\frac{l!}{(2l+1)!!}g^{l,m}_{i_{1} \cdots i_{l}}a^{+}_{l,m}   . 
\end{eqnarray}
\begin{equation}
	f^{l,m}_{i_{1} \cdots i_{l}} = \frac{l!}{(2l+1)!!}(g^{l,m}_{i_{1} \cdots i_{l}})^{*}. 
\end{equation}
As a result, 
\begin{equation}\label{5r6}
	[a^{k}_{i_{1}\cdots i_{l}}, a^{+k'}_{j_{1}\cdots j_{l}}]=[\frac{(2l+1)!!}{l!}]^{2}N_{i_{1}\cdots i_{l};\;j_{1}\cdots j_{l}}\delta_{k,k'}, 
\end{equation}
which is the equation we will frequently use in later calculations.

\section{The conformal group generators in $3d$ CFT}

For the $3d$ free scalar field theory, the traceless totally symmetric conserved stress tensor is
\begin{equation}
T^{\mu\nu}=\frac{1}{4}     [\frac{3}{2}(\partial^{\mu}\phi\partial^{\nu}\phi+\partial^{\nu}\phi \partial^{\mu}\phi)-\delta^{\mu \nu}\partial^{\alpha}\phi \partial_{\alpha}\phi-\frac{1}{2}(\phi \partial^{\mu}\partial^{\nu}\phi+ \partial^{\mu}\partial^{\nu}\phi \phi)  ].
\end{equation}
The $SO(3,2)$ conformal group is generated by operators $\{D,P^{\mu},K^{\mu},M^{\mu\nu}\}$. In radial quantization, these operators are given by \cite{1q}
\begin{equation}
	D = i \int_{S^{2}} d\Omega \;n_{\mu}n_{\nu} 	T^{\mu\nu}(1, \theta,\varphi)
\end{equation}
\begin{equation}
	P^{\mu}=i \int_{S^{2}} d\Omega \;n_{\nu} 	T^{\mu\nu}(1, \theta,\varphi)
\end{equation}
\begin{equation}
K^{\mu}=i \int_{S^{2}} d\Omega \;[2 n^{\mu}n_{\rho}n_{\sigma} 	T^{\rho\sigma}(1, \theta,\varphi)-n_{\nu}	T^{\mu\nu}(1, \theta,\varphi)]
\end{equation}
\begin{equation}
M^{\mu\nu}=i \int_{S^{2}} d\Omega \;[ n^{\mu}n_{\sigma} 	T^{\nu \sigma}(1, \theta,\varphi)-n^{\nu}n_{\sigma} 	T^{\mu \sigma}(1, \theta,\varphi)],	
\end{equation}
satisfying
\begin{equation}
	D^{+}=-D,\;\;\;\;\; (P^{\mu})^{+}=-K^{\mu},\;\;\;\;\; (K^{\mu})^{+}=-P^{\mu},\;\;\;\;\;(M^{\mu\nu})^{+}=M^{\mu\nu}.
\end{equation}
The Hamilton operator is $H=iD$.

As shown in Appendix A, with (\ref{rf}) plugged in, we obtain
\begin{eqnarray}
D \nonumber &=& -i \sum \;  \frac{N_{j_{1} \cdots j_{l};\;i_{1} \cdots i_{l}}}{4} [ (2l+1)(a^{+}_{j_{1}\cdots j_{l}}a_{i_{1}\cdots i_{l}}+  a_{j_{1}\cdots j_{l}}a^{+}_{i_{1}\cdots i_{l}})] \\\nonumber &=& - i \sum_{l,m} \;  \frac{1}{4} [ (2l+1)(a^{+}_{l,m}a_{l,m}+  a_{l,m}a^{+}_{l,m})] \\\nonumber &=& - i \sum_{l,m} \;   (l+\frac{1}{2}) a^{+}_{l,m}a_{l,m} + \textnormal{const} \\ &=&-  i \sum \;   (l+\frac{1}{2}) N_{j_{1} \cdots j_{l};\;i_{1} \cdots i_{l}} a^{+}_{j_{1}\cdots j_{l}}a_{i_{1}\cdots i_{l}} + \textnormal{const},
\end{eqnarray}
\begin{equation}
P_{\mu}= -i 	\sum \; \sqrt{(2l+1)(2l+3)} N_{j_{1}\cdots j_{l+1};\;\mu i_{1}\cdots i_{l}}a^{+}_{j_{1}\cdots j_{l+1}}a_{i_{1}\cdots i_{l}}. 
\end{equation}
Taking into account of the fact that $a^{+}$ and $a$ are traceless, we have
\begin{equation}\label{r567}
	D=-  i \sum \;   \frac{l!}{2(2l-1)!!} a^{+}_{i_{1}\cdots i_{l}}a_{i_{1}\cdots i_{l}} + \textnormal{const},
\end{equation}
\begin{equation}
P_{\mu}= -i 	\sum \;  \sqrt{\frac{2l+1}{2l+3}}  \frac{(l+1)!}{(2l+1)!!}  a^{+}_{\mu i_{1}\cdots i_{l}}a_{i_{1}\cdots i_{l}}, 	
\end{equation}
\begin{equation}
K_{\mu}= -i 	\sum \;  \sqrt{\frac{2l+1}{2l+3}}  \frac{(l+1)!}{(2l+1)!!}  a^{+}_{ i_{1}\cdots i_{l}}a_{\mu i_{1}\cdots i_{l}}, 	
\end{equation}
\begin{equation}\label{r567y}
	M_{\mu\nu} = i 	\sum \;  \frac{(l+1)(l+1)!}{2(2l+3)!!}( a^{+}_{\mu q_{1}\cdots q_{l}}a_{\nu q_{1}\cdots q_{l}}-a^{+}_{\nu q_{1}\cdots q_{l}}a_{\mu q_{1}\cdots q_{l}}),
\end{equation}
\begin{equation}
	H=iD= \sum \;   \frac{l!}{2(2l-1)!!} a^{+}_{i_{1}\cdots i_{l}}a_{i_{1}\cdots i_{l}} + \textnormal{const},
\end{equation}
which satisfy
\begin{equation}
	[H, a^{+}_{l,m}]= (l+\frac{1}{2})a^{+}_{l,m}, \;\;\;\;\;\;\;\;\;\;\;\;\;\;\;\;[H, a_{l,m}]= -(l+\frac{1}{2})a_{l,m},
\end{equation}
\begin{equation}
	[H, a^{+}_{i_{1}\cdots i_{l}}]= (l+\frac{1}{2})a^{+}_{i_{1}\cdots i_{l}}, \;\;\;\;\;\;\;\;\;\;\;\;\;\;\;\;[H, a_{i_{1}\cdots i_{l}}]= -(l+\frac{1}{2})a_{i_{1}\cdots i_{l}}.
\end{equation}
One can see with the commutator (\ref{5r6}), the operators in (\ref{r567})-(\ref{r567y}) do form the $so(3,2)$ algebra. Especially, $[H,P_{\mu}]=P_{\mu}$, $[H,K_{\mu}]=-K_{\mu}$, $P_{\mu}$ and $K_{\mu}$ increases and decreases the energy by $1$,
respectively.  Also, we have
\begin{equation}
	[P_{\mu}, a^{+}_{i_{1}\cdots i_{l}}]=-i \sqrt{\frac{2l+1}{2l+3}} (l+1) a^{+}_{\mu i_{1}\cdots i_{l}} ,
\end{equation}
\begin{equation}
	[P_{\mu}, a_{i_{1}\cdots i_{l}}]=i\sqrt{\frac{2l+1}{2l-1}}[\frac{2l-1}{l}(\delta_{\mu i_{1}}a_{i_{2}\cdots i_{l}}+\cdots)-\frac{2}{l}(\delta_{i_{1}i_{2}}a_{\mu i_{3}\cdots i_{l}}+\cdots)]. 
\end{equation}
It is easy to show
\begin{equation}
	[P_{\mu},\phi]=-i\partial_{\mu}\phi,
\end{equation}
\begin{equation}
	[D,\phi]=-i(x^{\mu}\partial_{\mu}\phi+\frac{1}{2}\phi)
\end{equation}
as is required.

\section{The conserved charges in $3d$ CFT and the higher spin algebra}

The traceless totally symmetric conserved current for the $3d$ free scalar field theory 
was constructed as follows \cite{2eq}
\begin{equation}\label{122}
	J_{i_{1}\cdots i_{s}} = \sum^{s}_{k=0}\frac{(-1)^{k}(-\frac{1}{2})!(s-\frac{1}{2})!}{k! (k-\frac{1}{2})!(s-k)! (s-k-\frac{1}{2})!}\partial_{(i_{1}}\cdots \partial_{i_{k}}\phi  \partial_{i_{k+1}}\cdots \partial_{i_{s})}\phi -\textnormal{traces} 
\end{equation}
for each spin $s$. There is a one-to-one correspondence between the operator $J_{i_{1}\cdots i_{s}}$ in $3d$ CFT and the metric-like spin-s field $h_{i_{1}\cdots i_{s}}$ in $AdS_{4}$.

Because $\partial^{\mu}J_{\mu i_{1}\cdots  i_{s-1}}=0$, 
\begin{equation}
	Q_{i_{1}\cdots i_{s-1} } = \int_{\Sigma} dS^{\mu} \; J_{\mu i_{1}\cdots  i_{s-1}}
\end{equation}
is a conserved charge. $J_{i_{1}\cdots i_{s}}$ contracting with the conformal killing tensors also gives the conserved current. In \cite{3qr}, it was shown that the conserved currents and charges in the $d$ dimensional free massless scalar field theory are described by various traceless two-row rectangular Young tableaux of the conformal algebra $so(d,2)$. Killing tensors could be built as the products of the conformal killing vectors satisfying 
\begin{equation}
	\partial^{\mu}f^{\nu}+\partial^{\nu}f^{\mu}-\frac{2}{3}\delta^{\mu\nu}\partial_{\lambda}f^{\lambda}=0. 
\end{equation}
In $3d$, there are three kinds of such non-constant killing vectors
\begin{equation}\label{killi}
k^{\mu}=	x^{\mu},\;\;\;\;\;\;\;\;k^{\mu;\alpha}=2x^{\mu}x^{\alpha}- \delta^{\mu \alpha}r^{2},\;\;\;\;\;\;\;\;k^{\mu;\alpha\beta}=\delta^{\mu \alpha}x^{\beta}-\delta^{\mu \beta}x^{\alpha}.
\end{equation}
So, generically, the conserved currents can be taken as
\begin{equation}\label{dr5g}
J_{\mu i_{1}\cdots  i_{s-1}},\;\;\;\;\;\;\;\; f^{i_{1}}_{1}J_{\mu i_{1}\cdots  i_{s-1}}	,\;\;\;\;\;\;\;\;f^{i_{1}}_{1}f^{i_{2}}_{2}J_{\mu i_{1}\cdots  i_{s-1}},\;\;\;\;\;\;\;\; f^{i_{1}}_{1}\cdots f^{i_{s-1}}_{s-1}J_{\mu i_{1}\cdots  i_{s-1}}, 
\end{equation}
where $f^{i_{p}}_{p}$ can be any of the three killing vectors in (\ref{killi}). For example, when $s=2$, the conserved currents are 
\begin{equation}
J_{\mu i_{1}},\;\;\;\;\;\;\;\; k^{i_{1}} J_{\mu i_{1}},\;\;\;\;\;\;\;\; k^{i_{1};\alpha} J_{\mu i_{1}},\;\;\;\;\;\;\;\;
k^{i_{1};\alpha\beta} J_{\mu i_{1}}; 
\end{equation}
when $s=3$, neglecting the fact that $J=0$ for real $\phi$, the conserved currents are 
\begin{eqnarray}
\nonumber && J_{\mu i_{1}i_{2}},\\\nonumber  && k^{i_{1}} J_{\mu i_{1}i_{2}},\;\;\;\;\;\;\;\;  k^{i_{1};\alpha} J_{\mu i_{1}i_{2}}, \;\;\;\;\;\;\;\; k^{i_{1};\alpha\beta} J_{\mu i_{1}i_{2}},\\\nonumber  && k^{i_{1}}k^{i_{2}} J_{\mu i_{1}i_{2}},\;\;\;\;\;\;\;\;  k^{i_{1};\alpha}k^{i_{2};\beta} J_{\mu i_{1}i_{2}}, \;\;\;\;\;\;\;\; k^{i_{1};\alpha\beta} k^{i_{2};\rho\sigma} J_{\mu i_{1}i_{2}},\\  && k^{i_{1}}k^{i_{2};\alpha} J_{\mu i_{1}i_{2}},\;\;\;\;\;\;\;\;  k^{i_{1};\alpha}k^{i_{2};\rho\sigma} J_{\mu i_{1}i_{2}}, \;\;\;\;\;\;\;\; k^{i_{1};\alpha\beta}k^{i_{2}} J_{\mu i_{1}i_{2}}.
\end{eqnarray}
The integration of conserved current gives the conserved charge. It is not quite obvious that all charges form a closed Lie algebra. However, let $Q$ be an arbitrary conserved charge, there will be 
\begin{equation}\label{f5r}
	[Q,\phi]=\mathcal{D}\phi,
\end{equation}
where $\mathcal{D}$ is a linear differential operator satisfying $\Delta \mathcal{D} = \delta \Delta$ for some linear differential operator $\delta$ so that for $\Delta \phi = 0$, $\Delta [Q,\phi] = 0$. Higher spin algebra is the algebra for $Q$, which is isomorphic to the algebra for $\mathcal{D}$, while the latter is shown to be the quotient of the universal enveloping algebra $\mathcal{U}(so(d,2))$ \cite{4ft} and then is of course closed.

In this Section, we will compute the charge $Q$ explicitly, which gives another oscillator realization of the higher spin algebra.  
\begin{equation}
	[D,J_{i_{1}\cdots  i_{s}}]=-i[x^{\mu}\partial_{\mu}J_{i_{1}\cdots  i_{s}}+ (s+1)J_{i_{1}\cdots  i_{s}}]. 
\end{equation}
$J_{i_{1}\cdots  i_{s}}$ has the conformal dimension $s+1$. $k^{\mu}$, $k^{\mu;\alpha}$ and $k^{\mu;\alpha\beta}$ have the conformal dimensions $-1$, $-2$ and $-1$, respectively, so
\begin{equation}
	[D,(k)^{l} (k^{\alpha})^{m}( k^{\alpha\beta})^{n} J]=-i[x^{\mu}\partial_{\mu}((k)^{l} (k^{\alpha})^{m}( k^{\alpha\beta})^{n} J)+ (s+1-l-2m-n)((k)^{l} (k^{\alpha})^{m}( k^{\alpha\beta})^{n} J)].  
\end{equation}
For any operator $j$ with
\begin{equation}
	[D,j]=-i[x^{\mu}\partial_{\mu}j+ \Delta j],  
\end{equation}
\begin{equation}
	[D,r^{2}d\Omega j]=-i[x^{\mu}\partial_{\mu}( r^{2}d\Omega  j)+ (\Delta-2)r^{2}d\Omega j],  
\end{equation}
so if $j$ is the conserved current, i.e. 
\begin{equation}
	Q = \int_{S^{2}} r^{2}d\Omega j
\end{equation}
is a constant, there will be 
\begin{equation}
	[D,Q]=[D,\int_{S^{2}}  r^{2}d\Omega j]=-i[x^{\mu}\partial_{\mu}(\int_{S^{2}}  r^{2}d\Omega  j)+ (\Delta-2)\int_{S^{2}} r^{2}d\Omega j] = -i(\Delta-2)Q.  
\end{equation}
$(k)^{l} (k^{\alpha})^{m}( k^{\alpha\beta})^{n} J$ is conserved with 
\begin{equation}
	Q(l,m,n,s)=\int_{S^{2}}r^{2} d\Omega\; (k)^{l} (k^{\alpha})^{m}( k^{\alpha\beta})^{n} J
\end{equation}
the corresponding charge, so
\begin{equation}
	[D,Q(l,m,n,s)]=-i (s-1-l-2m-n)Q(l,m,n,s).   
\end{equation}
$Q(l,m,n,s)$ has the definite conformal dimension.

Charges can only be the linear combinations of the terms like 
\begin{eqnarray}
\nonumber && a^{+}_{\nu_{1}\cdots  \nu_{\alpha} q_{1}\cdots  q_{l}}a_{\mu_{1}\cdots  \mu_{\beta}   q_{1}\cdots  q_{l}},\;\;\;\;\;\;\;\;\;\;a^{+}_{\nu_{1}\cdots  \nu_{\alpha} q_{1}\cdots  q_{l}} a^{+}_{\mu_{1}\cdots  \mu_{\beta}   q_{1}\cdots  q_{l}},\\\nonumber && a_{\nu_{1}\cdots  \nu_{\alpha} q_{1}\cdots  q_{l}}a_{\mu_{1}\cdots  \mu_{\beta}   q_{1}\cdots  q_{l}},\;\;\;\;\;\;\;\;\;\; a_{\nu_{1}\cdots  \nu_{\alpha} q_{1}\cdots  q_{l}}a^{+}_{\mu_{1}\cdots  \mu_{\beta}   q_{1}\cdots  q_{l}}.
\end{eqnarray}
Obviously, it is $a^{+}_{\nu_{1}\cdots  \nu_{\alpha} q_{1}\cdots  q_{l}}$ $a_{\mu_{1}\cdots  \mu_{\beta}   q_{1}\cdots  q_{l}}$ and $a_{\nu_{1}\cdots  \nu_{\alpha} q_{1}\cdots  q_{l}}a^{+}_{\mu_{1}\cdots  \mu_{\beta}   q_{1}\cdots  q_{l}}$ that are allowed with $\alpha-\beta=s-1-l-2m-n$ and $\beta-\alpha=s-1-l-2m-n$, respectively. Especially, the charge for $J_{\mu i_{1}\cdots  i_{s-1}}$ is the linear combination of $a^{+}_{i_{1}\cdots  i_{s-1} q_{1}\cdots  q_{l}}a_{q_{1}\cdots  q_{l}}$. With $J_{\mu i_{1}\cdots  i_{s-1}}\rightarrow k^{i_{1}}J_{\mu i_{1}\cdots  i_{s-1}}$, 
\begin{equation}
a^{+}_{i_{1}\cdots  i_{s-1} q_{1}\cdots  q_{l}}a_{q_{1}\cdots  q_{l}}\rightarrow 	a^{+}_{i_{2}\cdots  i_{s-1} q_{1}\cdots  q_{l}}a_{q_{1}\cdots  q_{l}};
\end{equation}
with $J_{\mu i_{1}\cdots  i_{s-1}}\rightarrow k^{i_{1};\alpha}J_{\mu i_{1}\cdots  i_{s-1}}$, 
\begin{equation}
a^{+}_{i_{1}\cdots  i_{s-1} q_{1}\cdots  q_{l}}a_{q_{1}\cdots  q_{l}}\rightarrow 	a^{+}_{i_{2}\cdots  i_{s-1} q_{1}\cdots  q_{l}}a_{\alpha q_{1}\cdots  q_{l}};
\end{equation}
with $J_{\mu i_{1}\cdots  i_{s-1}}\rightarrow k^{i_{1};\alpha\beta}J_{\mu i_{1}\cdots  i_{s-1}}$, 
\begin{equation}
a^{+}_{i_{1}\cdots  i_{s-1} q_{1}\cdots  q_{l}}a_{q_{1}\cdots  q_{l}}\rightarrow 	a^{+}_{\alpha i_{2}\cdots  i_{s-1} q_{1}\cdots  q_{l}}a_{\beta q_{1}\cdots  q_{l}}-a^{+}_{\beta i_{2}\cdots  i_{s-1} q_{1}\cdots  q_{l}}a_{\alpha q_{1}\cdots  q_{l}}. 
\end{equation}
In short, the action of $k^{\mu}$ is to eliminate the $\mu$ index from $a^{+}$; the action of $k^{\mu;\alpha}$ is to eliminate the $\mu$ index from $a^{+}$ and add the $\alpha$ index in $a$; the action of $k^{\mu;\alpha\beta}$ is to replace the $\mu$ index in $a^{+}$ by the $\alpha/\beta$ index and add the $\beta/\alpha$ index in $a$ with $\alpha$ and $\beta$ anti-symmetric. It is easy to see the constructed charge indeed has the expected conformal dimension.

In the following, we will calculate two special sets of charges $P^{s}_{\mu_{1}\cdots \mu_{k}}$ and $K^{s}_{\mu_{1}\cdots \mu_{k}}$, which are the higher spin generalization of $P_{\mu}$, $D$ and $K_{\mu}$.  
\begin{equation}\label{121}
	P^{s}_{\mu_{1}\cdots \mu_{k}}=i \int_{S^{2}} d\Omega \;k^{\mu_{k+1}}\cdots k^{\mu_{s}}	J_{\mu_{1}\cdots \mu_{s}};
\end{equation}
\begin{equation}
	K^{s}_{\mu_{1}\cdots \mu_{k}}=i \int_{S^{2}} d\Omega \;k^{\nu_{1};\mu_{1}}\cdots k^{\nu_{k};\mu_{k}}k^{\nu_{k+1}}\cdots k^{\nu_{s}}	J_{\nu_{1}\cdots \nu_{s}}-\textnormal{traces}, 
\end{equation}
\begin{equation}
	[D,P^{s}_{\mu_{1}\cdots \mu_{k}}]=-ikP^{s}_{\mu_{1}\cdots \mu_{k}},\;\;\;\;\;\;\;[D,K^{s}_{\mu_{1}\cdots \mu_{k}}]=i kK^{s}_{\mu_{1}\cdots \mu_{k}},
\end{equation}
where $K^{s}_{\mu_{1}\cdots \mu_{k}}=-(P^{s}_{\mu_{1}\cdots \mu_{k}})^{+}$.

The terms of interest are
\begin{eqnarray}\label{12}
\nonumber  && 	\partial_{\mu_{1}} \cdots\partial_{\mu_{n}}\phi^{+} (r,\theta,\varphi)\partial_{\nu_{1}} \cdots\partial_{\nu_{m}}\phi^{-} (r,\theta,\varphi)|_{r=1}  \\  &=& 
 \sum	\sum \; \frac{(-1)^{m}(l+n)! (2l'+2m-1)!!a^{+}_{\mu_{1}\cdots \mu_{n}i_{1} \cdots i_{l}} a_{j_{1} \cdots j_{l'}}}{l!(2l'-1)!!\sqrt{2l+2n+1}\sqrt{2l'+1}} Y_{i_{1}\cdots i_{l}} Y_{\nu_{1}\cdots \nu_{m}j_{1}\cdots j_{l'} } 
\end{eqnarray}
Plugging (\ref{12}) into (\ref{122}) and then (\ref{121}), one can see the generic form of $P^{s}_{\mu_{1}\cdots \mu_{k}}$ is
\begin{equation}\label{jio}
P^{s}_{\mu_{1}\cdots \mu_{k}}= -i 	\sum \; \frac{(l+k)!g(s,k,l)}{\sqrt{(2l+1)(2l+2k+1)} (2l+2k+1)!!} a^{+}_{\mu_{1}\cdots \mu_{k} i_{1}\cdots i_{l}}a_{i_{1}\cdots i_{l}}  
\end{equation}
with $g(s,k,l)$ a $s$-order polynomial of $l$. Let 
\begin{equation}
	f(s,k,l)=\frac{(2l-1)!!}{(2l+2k+1)!!}g(s,k,l), 
\end{equation}
(\ref{jio}) can be rewritten as 
\begin{equation}\label{rt5}
P^{s}_{\mu_{1}\cdots \mu_{k}}= -i 	\sum \; \sqrt{\frac{2l+1}{2l+2k+1}}\frac{(l+k)!f(s,k,l)}{ (2l+1)!!} a^{+}_{\mu_{1}\cdots \mu_{k} i_{1}\cdots i_{l}}a_{i_{1}\cdots i_{l}}.  
\end{equation}
\begin{eqnarray}
	[P^{s_{1}}_{\nu_{1}\cdots \nu_{k_{1}}}, P^{s_{2}}_{\mu_{1}\cdots \mu_{k_{2}}}] \nonumber &=& 	 \sum  \sqrt{\frac{2l+1}{2l+2k_{1}+2k_{2}+1}}  \frac{(l+k_{1}+k_{2})!\;a^{+}_{\nu_{1}\cdots \nu_{k_{1}} \mu_{1}\cdots \mu_{k_{2}} i_{1}\cdots i_{l}}a_{i_{1}\cdots i_{l}}}{(2l+1)!!} \\\nonumber &&  [f(s_{1},k_{1},l)f(s_{2},k_{2},l+k_{1})-f(s_{2},k_{2},l)f(s_{1},k_{1},l+k_{2})].  
\end{eqnarray}
The checking with the lower spin case shows that $f(s,k,l)$ is a $s-k-1$ order polynomial, so we will have
\begin{equation}\label{34e}
f(s_{1},k_{1},l)f(s_{2},k_{2},l+k_{1})-f(s_{2},k_{2},l)f(s_{1},k_{1},l+k_{2})= \sum^{s_{1}+s_{2}-2}_{s=k_{1}+k_{2}+1} \alpha^{s_{1},k_{1};s_{2},k_{2}}_{s} f(s,k_{1}+k_{2},l)	
\end{equation}
with $\alpha^{s_{1},k_{1};s_{2},k_{2}}_{s}$ uniquely determined. As a result, 
\begin{equation}
	[P^{s_{1}}_{\nu_{1}\cdots \nu_{k_{1}}}, P^{s_{2}}_{\mu_{1}\cdots \mu_{k_{2}}}]=i\sum^{s_{1}+s_{2}-2}_{s=k_{1}+k_{2}+1} \alpha^{s_{1},k_{1};s_{2},k_{2}}_{s} P^{s}_{\nu_{1}\cdots \nu_{k_{1}}\mu_{1}\cdots \mu_{k_{2}}}. 
\end{equation}
Obviously, $[P^{s},P^{s'}]=0$, $[P^{s}_{\nu_{1}\cdots \nu_{s-1}}, P^{s'}_{\nu_{1}\cdots \nu_{s'-1}}]=0$, $[P^{s}_{\nu_{1}\cdots \nu_{k}}, P^{s}_{\mu_{1}\cdots \mu_{k}}]=0$.

$f(s,0,l)$ can be calculated as
\begin{equation}
	f(s,0,l)=\frac{(-1)^{s}}{2l+1}\sum^{s}_{k=0} C^{2k}_{2s}C^{s}_{l+s-k}. 
\end{equation}
$f(2,1,l)=4$. When $s_{2}=2$, $k_{2}=1$, 
\begin{equation}
	[P^{s_{1}}_{\nu_{1}\cdots \nu_{k_{1}}}, P^{2}_{\mu_{1}}]=i\sum^{s_{1}}_{s=k_{1}+2} \alpha^{s_{1},k_{1};2,1}_{s} P^{s}_{\nu_{1}\cdots \nu_{k_{1}}\mu_{1}}. 
\end{equation}
$P^{s_{1}}_{\nu_{1}\cdots \nu_{k_{1}}}$ is a tensor in the representation of $so(3,2)$, so $	[P^{s_{1}}_{\nu_{1}\cdots \nu_{k_{1}}}, P^{2}_{\mu_{1}}]$ should also be a spin-$s$ operator. We actually have 
\begin{equation}
	[P^{s_{1}}_{\nu_{1}\cdots \nu_{k_{1}}}, P^{2}_{\mu_{1}}]=i \alpha^{s_{1},k_{1};2,1}_{s_{1}} P^{s_{1}}_{\nu_{1}\cdots \nu_{k_{1}}\mu_{1}}, 
\end{equation}
\begin{equation}
4f(s_{1},k_{1},l)-4f(s_{1},k_{1},l+1)=  \alpha^{s_{1},k_{1};2,1}_{s_{1}} f(s,k_{1}+1,l). 	
\end{equation}
From $f(s,0,l)$, $f(s,k,l)$ for $k=1,\cdots,s-1$ can be determined as 
\begin{equation}
	f(s,1,l)= \frac{f(s,0,l+1)-f(s,0,l)}{s-1},
\end{equation}
$\;\;\;\;\;\;\;\;\;\;\;\;\;\;\;\;\;\;\;\;\;\;\;\;\;\;\;\;\;\;\;\;\;\;\;\;\;\;\;\;\;\;\;\;\;\;\;\;\;\;\;\;\;\;\;\;\;\;\;\;\;\;\;\; \cdots$
\begin{equation}
	f(s,k,l)= \frac{f(s,k-1,l+1)-f(s,k-1,l)}{s-k},
\end{equation}
$\;\;\;\;\;\;\;\;\;\;\;\;\;\;\;\;\;\;\;\;\;\;\;\;\;\;\;\;\;\;\;\;\;\;\;\;\;\;\;\;\;\;\;\;\;\;\;\;\;\;\;\;\;\;\;\;\;\;\;\;\;\;\;\;\cdots$
\begin{equation}
	f(s,s-1,l)= f(s,s-2,l+1)-f(s,s-2,l). 
\end{equation}

For $s=2,3,4$, the direct calculation gives
\begin{eqnarray}
\nonumber &&	f(2,0,l)=4l+2,\;\;\;\;\;\;	f(2,1,l)=4; \\\nonumber && 	f(3,0,l)=\frac{2}{3}(8 l^2+8 l+3),\;\;\;\;\;\;	f(3,1,l)=\frac{16}{3}( l+1 ),\;\;\;\;\;\;	f(3,2,l)=\frac{16}{3};  \\\nonumber &&  f(4,0,l)=\frac{1}{3}(8 l^3+12 l^2+10 l+3),\;\;\;\;\;\;	f(4,1,l)=\frac{2}{3}(4 l^2+8 l+5),\\\nonumber && 	f(4,2,l)=\frac{4}{3}(2l+3),\;\;\;\;\;\;	f(4,3,l)=\frac{8}{3};  
\end{eqnarray}
verifying our expectation. The algebra is then
\begin{eqnarray}
\nonumber &&		[P^{2}_{\mu},P^{3}]\propto P^{3}_{\mu},\;\;\;\;\;[P^{2}_{\mu},P^{3}_{\nu}]\propto P^{3}_{\mu\nu}; \\\nonumber && 	[P^{2}_{\mu},P^{4}_{\nu\lambda}]\propto P^{4}_{\mu\nu\lambda},\;\;\;\;\;[P^{2}_{\mu},P^{4}]\propto P^{4}_{\mu} ,\;\;\;\;\;[P^{2}_{\mu},P^{4}_{\nu}]\propto P^{4}_{\mu\nu};  \\\nonumber &&  [P^{3},P^{3}_{\mu}]\propto 12 P^{4}_{\mu}-\frac{1}{2}P^{2}_{\mu},\;\;\;\;\;[P^{3},P^{3}_{\mu\nu}]\propto P^{4}_{\mu\nu}.  
\end{eqnarray}

With $P^{s}_{\mu_{1}\cdots \mu_{k}}$ given, since $K^{s}_{\mu_{1}\cdots \mu_{k}}=-(P^{s}_{\mu_{1}\cdots \mu_{k}})^{+}$, 
\begin{equation}
K^{s}_{\mu_{1}\cdots \mu_{k}}= -i 	\sum \; \sqrt{\frac{2l+1}{2l+2k+1}}\frac{(l+k)!f(s,k,l)}{ (2l+1)!!} a^{+}_{i_{1}\cdots i_{l}}a_{\mu_{1}\cdots \mu_{k} i_{1}\cdots i_{l}}.  
\end{equation}
$\{P^{s}_{\mu_{1}\cdots \mu_{k}}\}$ and $\{K^{s}_{\mu_{1}\cdots \mu_{k}}\}$ form two subalgebras of the higher spin algebra. With some tedious calculations, the rest higher spin charges can also be worked out, giving rise to the CFT realization of the higher spin charges which is supposed to be constructed from the higher spin gauge fields in $AdS_{4}$. In radial quantization, the charges listed in (\ref{dr5g}) have the definite conformal dimension but are not necessarily Hermitian. If $Q$ is a conserved charge with the dimension $\Delta$, $Q^{+}$ will also be the conserved charge with the dimension $-\Delta$. We will then get the Hermitian charges $Q+Q^{+}$ and $iQ-iQ^{+}$ that does not have the definite conformal dimension.

\begin{equation}
	[P^{s}_{\mu_{1}\cdots \mu_{k}}, a^{+}_{i_{1}\cdots i_{l}}]= - i	\sqrt{\frac{2l+1}{2l+2k+1}} \frac{(l+k)!f(s,k,l)}{ l!} a^{+}_{\mu_{1}\cdots \mu_{k} i_{1}\cdots i_{l}}.
\end{equation}
Especialy, 
\begin{equation}
	[P^{s}_{\mu_{1}\cdots \mu_{s-1}}, a^{+}_{i_{1}\cdots i_{l}}] \propto - i	\sqrt{\frac{2l+1}{2l+2s-1}} \frac{(l+s-1)! }{ l!} a^{+}_{\mu_{1}\cdots \mu_{s-1} i_{1}\cdots i_{l}}, 
\end{equation}
\begin{equation}
	[P^{s}_{\mu_{1}\cdots \mu_{s-1}}, \phi] \propto -i \partial_{\mu_{1}}\cdots \partial_{\mu_{s-1}}\phi. 
\end{equation}
More generically,
\begin{equation}
	[P^{s}_{\mu_{1}\cdots \mu_{k}}, \phi] \propto -i \sum^{s-1}_{n=k}[V_{s,k,n}\; x^{\mu_{k+1}}\cdots x^{\mu_{n}}	 \partial_{\mu_{1}}\cdots \partial_{\mu_{n}}\phi], 
\end{equation}
where $V_{s,k,n}$ are coefficients to be determined via
\begin{equation}
	f(s,k,l)=\sum^{s-1}_{n=k}\;[\frac{l!}{(l+k-n)!}V_{s,k,n}].
\end{equation}
Because
\begin{equation}
\Delta(x^{\mu_{k+1}}\cdots x^{\mu_{n}}	 \partial_{\mu_{1}}\cdots \partial_{\mu_{n}}\phi)=0,
\end{equation}
\begin{equation}
\Delta	[P^{s}_{\mu_{1}\cdots \mu_{k}}, \phi]=0, 
\end{equation}
is consistent with (\ref{f5r}).

In \cite{3qr}, it was shown that the higher spin algebra $ho(1|2:[3,2])$ admits a basis formed by a set of elements $M_{A_{1}\cdots A_{s-1},B_{1}\cdots B_{s-1}}$ in irreducible representations of $SO(3,2)$ characterized by two row
rectangular Young tableaux. $A_{i},B_{i} = 0,1,2,3,4$. 
\begin{eqnarray}\label{t6y7u1}
\nonumber && 	M_{A_{1}\cdots A_{s-1},B_{1}\cdots B_{s-1}}=M_{\{A_{1}\cdots A_{s-1}\},B_{1}\cdots B_{s-1}}=M_{A_{1}\cdots A_{s-1},\{B_{1}\cdots B_{s-1}\}},  \\&& M_{\{A_{1}\cdots A_{s-1},A_{s}\}B_{2}\cdots B_{s-1}}=0,\;\;\;\;\;\;\;M_{A_{1}\cdots A_{s-3}CC,}^{\;\;\;\;\;\;\;\;\;\;\;\;\;\;\;\;\;\;\;\;\; B_{1}\cdots B_{s-1}}=0. 
\end{eqnarray}
When $s=2$, $\{M_{A_{1},B_{1}}\}$ gives the $so(3,2)$ algebra. All generators transform as $SO(3,2)$ tensors:
\begin{equation}\label{t6y7u}
	[M_{C,D}, M_{A_{1}\cdots A_{s-1},B_{1}\cdots B_{s-1}}]=i\eta_{DA_{1}}M_{CA_{2}\cdots A_{s-1},B_{1}\cdots B_{s-1}}+\cdots.
\end{equation}
We can make a mapping from $\{P^{s}_{\mu_{1}\cdots \mu_{k}}\}$ and $\{K^{s}_{\mu_{1}\cdots \mu_{k}}\}$ to $\{M_{A_{1}\cdots A_{s-1},B_{1}\cdots B_{s-1}}\}$. For $s=2$, 
\begin{equation}
	D^{2} =i M_{0,4}, \;\;\;\;\;\;\; P_{\mu_{1}}^{2} = M_{0,\mu_{1}}+iM_{4,\mu_{1}}, \;\;\;\;\;\;\; K_{\mu_{1}}^{2} = -M_{0,\mu_{1}}+iM_{4,\mu_{1}}. 
\end{equation}
According to (\ref{t6y7u1}), for each spin $s$, there is a unique $SO(3)$ scalar $M_{0\cdots 0,4\cdots 4}$,
\begin{equation}
	[M_{0,4},M_{0\cdots 0,4\cdots 4}]=0,
\end{equation}
so
\begin{equation}
	D^{s} = M_{0\cdots 0,4\cdots 4}, 
\end{equation}
$[D^{2},D^{s}]=0$. There are two $SO(3)$ vectors $M_{0\cdots 0,4\cdots 4\mu_{1}}$ and $M_{4\cdots 4,0\cdots 0\mu_{1}}$,
\begin{equation}
	[M_{0,4},M_{0\cdots 0,4\cdots 4\mu_{1}}]=i(-1)^{s-1} M_{4\cdots 4,0\cdots 0\mu_{1}},\;\;\;\;\;\;\;[M_{0,4},M_{4\cdots 4,0\cdots 0\mu_{1}}]=i(-1)^{s}M_{0\cdots 0,4\cdots 4\mu_{1}}. 
\end{equation}
\begin{equation}
	P^{s}_{\mu_{1}}=M_{0\cdots 0,4\cdots 4\mu_{1}}+i M_{4\cdots 4,0\cdots 0\mu_{1}},\;\;\;\;\;\;\;	K^{s}_{\mu_{1}}=-M_{0\cdots 0,4\cdots 4\mu_{1}}+i M_{4\cdots 4,0\cdots 0\mu_{1}},
\end{equation}
\begin{equation}
	[D^{2}, P^{s}_{\mu_{1}}]=i(-1)^{s-1}P^{s}_{\mu_{1}},\;\;\;\;\;\;\;\;\;\;[D^{2}, K^{s}_{\mu_{1}}]=i(-1)^{s}K^{s}_{\mu_{1}}. 
\end{equation}
For $k \geq 2$, the relation between $P^{s}_{\mu_{1} \cdots \mu_{k}}$, $K^{s}_{\mu_{1} \cdots \mu_{k}}$ and $M_{A_{1}\cdots A_{s-1},B_{1}\cdots B_{s-1}}$ is more complicated. Nevertheless, with $D = iM_{0,4}$, the higher spin generators $M_{A_{1}\cdots A_{s-1},B_{1}\cdots B_{s-1}}$ can be rearranged as the generators with the definite conformal weight $\Delta$, which will then correspond to the CFT operators taking the form of $\sum a^{+}_{\mu_{1}\cdots\mu_{\alpha}q_{1}\cdots q_{l}}a_{\nu_{1}\cdots\nu_{\beta}q_{1}\cdots q_{l}}$, with $\alpha-\beta=\Delta$. Especially, $[D, M_{\mu_{1}\cdots\mu_{k},\nu_{1}\cdots\nu_{k}}]=0$.

The above discussion can be directly extended to dimension $d$. The totally symmetric traceless conserved spin-s tensor is 
\begin{equation}
	J^{d}_{i_{1}\cdots i_{s}} = \sum^{s}_{k=0}\frac{(-1)^{k}(\frac{d}{2}-2)!(s+\frac{d}{2}-2)!}{k! (k+\frac{d}{2}-2)!(s-k)! (s-k+\frac{d}{2}-2)!}\partial_{(i_{1}}\cdots \partial_{i_{k}}\phi  \partial_{i_{k+1}}\cdots \partial_{i_{s})}\phi -\textnormal{traces}. 
\end{equation}
When $s=2$,
\begin{equation}
	J^{d}_{\mu\nu}= \frac{1}{2}(\phi \partial_{\mu} \partial_{\nu}\phi+ \partial_{\mu} \partial_{\nu} \phi \phi  )-\frac{d}{2(d-2)}(\partial_{\mu}\phi \partial_{\nu}\phi+\partial_{\nu}\phi \partial_{\mu}\phi -\frac{2}{d}\delta_{\mu\nu}\partial^{\alpha}\phi \partial_{\alpha}\phi),      
\end{equation}
from which, we get 
\begin{equation}
	D^{d} = -i\sum \frac{(d-1)l!}{(d-2)(2l+d-4)!!}a^{+}_{i_{1}\cdots i_{l}}a_{i_{1}\cdots i_{l}},
\end{equation}
\begin{equation}
	P^{d}_{\mu} =-i \sum \sqrt{\frac{2l+d-2}{2l+d}}\frac{2(d-1)(l+1)!}{(d-2)(2l+d-2)!!}a^{+}_{\mu i_{1}\cdots i_{l}}a_{i_{1}\cdots i_{l}}.
\end{equation}
The generic form of $P^{s,d}_{\mu_{1}\cdots \mu_{n}}$ is
\begin{equation}
P^{s,d}_{\mu_{1}\cdots \mu_{k}}= -i 	\sum \; \sqrt{\frac{2l+d-2}{2l+2k+d-2}}\frac{(l+k)!f^{d}(s,k,l)}{ (2l+d-2)!!} a^{+}_{\mu_{1}\cdots \mu_{k} i_{1}\cdots i_{l}}a_{i_{1}\cdots i_{l}},  
\end{equation}
with $f^{d}(s,k,l)$ a $s-k-1$-order polynomial.  
\begin{equation}
K^{s,d}_{\mu_{1}\cdots \mu_{k}}=-(P^{s,d}_{\mu_{1}\cdots \mu_{k}})^{+}= -i 	\sum \; \sqrt{\frac{2l+d-2}{2l+2k+d-2}}\frac{(l+k)!f^{d}(s,k,l)}{ (2l+d-2)!!} a^{+}_{i_{1}\cdots i_{l}} a_{\mu_{1}\cdots \mu_{k} i_{1}\cdots i_{l}}.  
\end{equation}

\section{The state correspondence for the higher spin/vector model duality}

With the radial quantization of the CFT done, we are ready to build the one-to-one correspondence between the Hilbert space of the higher spin gravity in AdS and the Hilbert space of the CFT. Consider the quantum higher spin gravity in $AdS_{4}$. The vacuum state could be denoted as $\left| \Omega \right\rangle$. $AdS_{4}$ has the $SO(3,2)$ isometry, so the operators $P_{\mu}$, $K_{\mu}$, $H$ and $M_{\mu\nu}$ with $\mu,\nu=1,2,3$ can be constructed from the theory in $AdS_{4}$. Moreover, higher spin gravity also has the higher spin symmetry, so the higher spin charges, including $P^{s}_{\mu_{1}\cdots\mu_{s}}$ and $K^{s}_{\mu_{1}\cdots\mu_{s}}$, can also be built in AdS theory. According to the AdS/CFT correspondence, these operators in AdS, although have the different origin, can be identified with the corresponding operators in CFT.

The one particle states of the higher spin gravity in $AdS_{4}$ form the infinite dimensional unitary reducible representation of $SO(3,2)$ \cite{5ft,6ft,7ft}. For each spin $s$, there is a lowest energy state $\left|\lambda, s \right\rangle$ with $K_{\mu} \left|\lambda, s \right\rangle=0$. The Hilbert space of the single particle state for spin-$s$ field is generated by $\{P_{\mu_{1}}\cdots P_{\mu_{n}} \left|\lambda, s \right\rangle|n=0,1\cdots\}$. 
\begin{equation}
	H\left|\lambda, s \right\rangle=\lambda\left|\lambda, s \right\rangle = (s+1)\left|\lambda, s \right\rangle.
\end{equation}
\begin{equation}
	H P_{\mu_{1}}\cdots P_{\mu_{n}} \left|\lambda, s \right\rangle=(\lambda+n)\left|\lambda, s \right\rangle =(s+n+1)\left|\lambda, s \right\rangle .
\end{equation}
$P_{\mu_{1}}\cdots P_{\mu_{n}} $ can be decomposed into the traceless tensors $\{P^{(n)}_{\mu_{1}\cdots \mu_{l}}|l=n,n-2,\cdots\}$, which are the unitary irreducible representations of $SO(3)$.   
\begin{eqnarray}
\nonumber && P^{(n)}_{\mu_{1}\cdots \mu_{n}} = P_{\mu_{1}}\cdots P_{\mu_{n}}-\textnormal{traces},  \\\nonumber&&  P^{(n)}_{\mu_{1}\cdots \mu_{n-2}} = P_{\mu_{1}}\cdots P_{\mu_{n-2}}P^{2}-\textnormal{traces},    \\\nonumber&&  P^{(n)}_{\mu_{1}\cdots \mu_{n-4}} = P_{\mu_{1}}\cdots P_{\mu_{n-4}}P^{4}-\textnormal{traces},\\\nonumber&&\cdots \\\nonumber&&	P^{(n)} = P^{n}\;\;\;\; \textnormal{for even}\;\; n,  \\&& P^{(n)}_{\mu_{1}} = P_{\mu_{1}}P^{n-1}\;\;\;\; \textnormal{for odd}\;\; n. 
\end{eqnarray}
$P^{(n)}_{\mu_{1}\cdots \mu_{l}} \left|\lambda, s \right\rangle =\left|n,l;\lambda, s \right\rangle$. Acting on the lowest energy state, $P^{(n)}_{\mu_{1}\cdots \mu_{l}}$ will give the orbit angular momentum $l$ and the external energy $n$.

The 1-particle Hilbert space of the spin-$s$ field have the basis $\{ \left|n,l;\lambda, s \right\rangle |n=0,1,\cdots;l=n,n-2\cdots\}$, forming the unitary irreducible representation of $SO(3,2)$. Let $C_{2}$ be the quadratic Casimir operator of $SO(3,2)$, then 
\begin{equation}
C_{2}	\left|n,l;\lambda, s \right\rangle =[\lambda(\lambda-3)+s(s+1)] \left|n,l;\lambda, s \right\rangle =2(s^{2}-1)  \left|n,l;\lambda, s \right\rangle.
\end{equation}
Under the $SO(3,2)$ transformation, both the energy and the angular momentum change, but the spin is preserved. The 1-particle Hilbert space for higher spin gravity in $AdS_{4}$ is then decomposed into the subspaces for each spin. Under the higher spin symmetry transformation, particles with the different spins will be mixed. Let $Q$ be a generator of the higher spin transformation, we may expect 
\begin{equation}
Q	\left|n,l;\lambda, s \right\rangle =\sum_{n',l',\lambda', s'}Q^{n,l,\lambda, s}_{n',l',\lambda', s'}\left|n',l';\lambda', s' \right\rangle. 
\end{equation}
The matrix $Q^{n,l,\lambda, s}_{n',l',\lambda', s'}$ gives the representation of the higher spin algebra in the 1-particle Hilbert space of the higher spin gravity, whose exact form can be determined in $3d$ CFT via the AdS/CFT correspondence.

Now, back to the CFT side. In $3d$ free scalar field theory, there is a set of spin-s primary operators  
\begin{equation}
	O_{i_{1}\cdots i_{s}} = \sum^{s}_{k=0}\frac{(-1)^{k}(-\frac{1}{2})!(s-\frac{1}{2})!}{k! (k-\frac{1}{2})!(s-k)! (s-k-\frac{1}{2})!}\partial_{\{i_{1}}\cdots \partial_{i_{k}}\phi  \partial_{i_{k+1}}\cdots \partial_{i_{s}\}}\phi -\textnormal{traces}. 
\end{equation}
\begin{equation}
	[H,O_{i_{1}\cdots i_{s}}(x)]=x^{\mu}\partial_{\mu}O_{i_{1}\cdots i_{s}}(x)+(s+1)O_{i_{1}\cdots i_{s}}(x). 
\end{equation}
Especially, 
\begin{equation}
	[H,O_{i_{1}\cdots i_{s}}(0)]=(s+1)O_{i_{1}\cdots i_{s}}(0),
\end{equation}
if $O_{i_{1}\cdots i_{s}}(x)$ is regular at $x=0$. 
\begin{equation}
	O_{i_{1}\cdots i_{s}}(x)=	O^{++}_{i_{1}\cdots i_{s}}(x)+O^{+-}_{i_{1}\cdots i_{s}}(x)+O^{-+}_{i_{1}\cdots i_{s}}(x)+O^{--}_{i_{1}\cdots i_{s}}(x),
\end{equation}
where $O^{++}$, $O^{+-}$, $O^{-+}$, $O^{--}$ are operators constructed from $\phi^{+}\phi^{+}$, $\phi^{+}\phi^{-}$, $\phi^{-}\phi^{+}$, $\phi^{-}\phi^{-}$ respectively. $O^{\alpha\beta}_{i_{1}\cdots i_{s}}(x)\sim \sum_{n,l} r^{n}Y_{l}$. For $O^{++}$, $n\geq l$; for $O^{+-}$, $O^{-+}$, $O^{--}$, $n < l$, so $O^{++}$ is the analytic part of $O$ at $x=0$. We will only keep the analytic part of $O_{i_{1}\cdots i_{s}}(x)$, for which, the descendants $\partial_{\mu_{1}}\cdots\partial_{\mu_{1}}O_{i_{1}\cdots i_{s}}(0)$ is 
well-defined~\footnote{Here, the analytic part is just the creation operator. For the generic operator $O(x)$ in an interacting CFT, the 
\begin{equation}
	O(x) = \sum_{n,l} r^{n}Y_{l} \;o_{n,l}
\end{equation}
decomposition is always possible, from which, the analytic part with $n\geq l$ can be uniquely determined and can also be taken as the creation operator.}. With $\phi^{+}$ plugged in, 
\begin{equation}\label{f5}
O_{i_{1}\cdots i_{s}}(0)=\sum^{s}_{k=0}\frac{(-1)^{k}(-\frac{1}{2})!(s-\frac{1}{2})!a^{+}_{\{i_{1}\cdots i_{k}}a^{+}_{i_{k+1}\cdots i_{s}\}}}{(k-\frac{1}{2})! (s-k-\frac{1}{2})!\sqrt{2k+1}\sqrt{2s-2k+1}}- \textnormal{traces}. 
\end{equation}

Suppose $\left| 0 \right\rangle$ is the vacuum state of the $3d$ CFT, $H \left| 0 \right\rangle=0$~\footnote{The vacuum energy in $H$ is dropped.}, then $O_{i_{1}\cdots i_{s}}(0)\left| 0 \right\rangle$ is a state with spin $s$, energy $s+1$. This is the standard operator/state correspondence in CFT. 
\begin{eqnarray}\label{curr}
\nonumber && O_{i_{1}\cdots i_{s}}(0)\left| 0 \right\rangle  \\ &=& \sum^{s}_{k=0}\frac{(-1)^{k}(-\frac{1}{2})!(s-\frac{1}{2})!}{\sqrt{(2k+1)(2s-2k+1)} (k-\frac{1}{2})! (s-k-\frac{1}{2})!}a^{+}_{\{i_{1}\cdots i_{k}}a^{+}_{i_{k+1}\cdots i_{s}\}} \left|0\right\rangle-\textnormal{traces}. 
\end{eqnarray}
For example, 
\begin{equation}
O(0)\left|0\right\rangle =a^{+}a^{+} 	\left|0\right\rangle, 
\end{equation}
\begin{equation}
	O_{i_{1}i_{2}}(0)\left|0\right\rangle =  (\frac{2}{\sqrt{5}}a^{+}a^{+}_{i_{1}i_{2}}-a^{+}_{i_{1}}a^{+}_{i_{2}}+\frac{1}{3}\delta_{i_{1}i_{2}}a^{+}_{\alpha}a^{+}_{\alpha})\left|0\right\rangle. 
\end{equation}
According to the AdS/CFT dictionary, the identification 
\begin{equation}
	O_{i_{1}\cdots i_{s}}(0)\left|0\right\rangle=\left|s+1,s\right\rangle
\end{equation}
can be made. $O(0)\left|0\right\rangle$ is a 1-particle state for the spin 0 field in $AdS_{4}$ with the energy $1$; $O_{i_{1}i_{2}}(0)\left|0\right\rangle$ is a 1-particle state for the spin 2 field in $AdS_{4}$ with the energy $3$. $O_{i_{1}\cdots i_{s}}$ is primary, $	[K_{\mu},O_{i_{1}\cdots i_{s}}]=0$, so $K_{\mu}	O_{i_{1}\cdots i_{s}}(0)\left|0\right\rangle=0$. $O_{i_{1}\cdots i_{s}}(0)\left|0\right\rangle$ is the lowest energy state. In fact, since 
\begin{equation}
	[K_{\mu}, a^{+}_{i_{1}\cdots i_{l}}]=-i\sqrt{\frac{2l+1}{2l-1}}[\frac{2l-1}{l}(\delta_{\mu i_{1}}a^{+}_{i_{2}\cdots i_{l}}+\cdots)-\frac{2}{l}(\delta_{i_{1}i_{2}}a^{+}_{\mu i_{3}\cdots i_{l}}+\cdots)],
\end{equation}
let $O_{i_{1}\cdots i_{s}}(0)\left|0\right\rangle$ be the most generic linear combination of the $a^{+}a^{+}\left|0\right\rangle$ terms with the energy $s+1$, i.e.  
\begin{equation}
O_{i_{1}\cdots i_{s}}(0) \left|0\right\rangle= (a^{+}_{i_{1}\cdots i_{s}}a^{+}+	f_{1}a^{+}_{i_{1}\cdots i_{s-1}}a^{+}_{i_{s}}+ f_{2}a^{+}_{i_{1}\cdots i_{s-2}i_{s}}a^{+}_{i_{s-1}}+ \cdots)\left|0\right\rangle, 
\end{equation}
imposing $K_{\mu}O_{i_{1}\cdots i_{s}}(0) \left|0\right\rangle=0$ will uniquely fix $O_{i_{1}\cdots i_{s}}(0) \left|0\right\rangle$ to be (\ref{curr}).

The boost operator $P_{\mu}$ has been constructed in CFT.  
\begin{equation}
	P_{\mu_{1}}\cdots P_{\mu_{n}} O_{i_{1}\cdots i_{s}}(0)\left| 0 \right\rangle= [P_{\mu_{1}},\cdots, [P_{\mu_{n-1}}, [P_{\mu_{n}}, O_{i_{1}\cdots i_{s}}(0)]] \cdots]\left| 0 \right\rangle. 
\end{equation}
It is direct to write down $P^{(n)}_{\mu_{1}\cdots \mu_{l}} \left|\lambda, s \right\rangle =\left|n,l;\lambda, s \right\rangle$. Let us give some simple examples. For the spin $0$ particle in $AdS_{4}$, the lowest energy state is $\left| 1, 0 \right\rangle =a^{+}a^{+} \left| 0 \right\rangle $. 
\begin{equation}
P^{(1)}_{\mu} \left| 1, 0 \right\rangle= P^{(1)}_{\mu}  a^{+} a^{+}\left| 0 \right\rangle = -\frac{2i}{\sqrt{3}}a_{\mu}^{+} a^{+}\left| 0 \right\rangle = \left|1,1; 1, 0 \right\rangle 
\end{equation}
is the state with the angular momentum $1$, energy $2$.  
\begin{equation}
P^{(2)}_{\mu \nu}  \left| 1,0 \right\rangle=P^{(2)}_{\mu \nu}  a^{+} a^{+}\left| 0 \right\rangle=-2(\frac{2}{\sqrt{5}}a_{\mu \nu}^{+} a^{+}+ \frac{1}{3}a_{\mu }^{+} a^{+}_{\nu}-\frac{\delta_{\mu \nu}}{9}a_{\alpha }^{+} a^{+}_{\alpha})\left| 0 \right\rangle =  \left|2,2; 1,0 \right\rangle 
\end{equation}
is the state with the angular momentum $2$, energy $3$.
\begin{equation}
	P^{(2)} \left| 1, 0 \right\rangle=a_{\alpha }^{+} a^{+}_{\alpha}\left| 0 \right\rangle=  \left|2,0; 1,0 \right\rangle
\end{equation}
is the state with the angular momentum $0$, energy $3$.

Since $[P_{\mu},O_{i_{1}\cdots i_{s}}]= -i \partial_{\mu} O_{i_{1}\cdots i_{s}}$, we actually have 
\begin{eqnarray}
\nonumber && P^{(n)}_{\mu_{1}\cdots \mu_{l}}  \left|s+1,s \right\rangle =P^{(n)}_{\mu_{1}\cdots \mu_{l}} O_{i_{1}\cdots i_{s}} (0)\left|0 \right\rangle \\\nonumber  &=& 	(-i)^{n}[ \partial_{\mu_{1}}\cdots \partial_{\mu_{l}}(\partial^{2})^{\frac{n-l}{2}} -\textnormal{traces}]O_{i_{1}\cdots i_{s}}(0) \left| 0 \right\rangle  \\  &=& O^{n}_{\mu_{1}\cdots \mu_{l};\;i_{1}\cdots i_{s}}(0) \left| 0 \right\rangle . 
\end{eqnarray}
The primary operator $O_{i_{1}\cdots i_{s}}$ corresponds to a state of spin s particle with the energy $s+1$, while the descendant operator $[ \partial_{\mu_{1}}\cdots \partial_{\mu_{l}}(\partial^{2})^{\frac{n-l}{2}} -\textnormal{traces}]O_{i_{1}\cdots i_{s}}$ corresponds to a state of the spin $s$ particle with the energy $n+s+1$ and the orbit angular momentum $l$ \cite{ft67y}. A special case is $P^{(n)}_{\mu_{1}\cdots \mu_{n}}$, 
\begin{equation}
P^{(n)}_{\mu_{1}\cdots \mu_{n}}  \left|s+1,s \right\rangle =	(-i)^{n}[\partial_{\mu_{1}}\cdots \partial_{\mu_{n}} -\textnormal{traces}]O_{i_{1}\cdots i_{s}}(0) \left| 0 \right\rangle.
\end{equation}
$[\partial_{\mu_{1}}\cdots \partial_{\mu_{n}} -\textnormal{traces}]$ could be compared with the chiral primary operator
\begin{equation}
	O_{I_{1}\cdots I_{n}} = X_{I_{1}}\cdots X_{I_{n}}-\textnormal{traces}
\end{equation}
in, for example, $N=4$ SYM theory. The former generates the angular momentum in $S^{p-2}$ of $AdS_{p}$, while the latter gives the angular momentum in the transverse $S^{q}$. In the longitudinal space, $X_{I}$ is replaced by $-i\partial_{\mu}$.

As is proposed in \cite{10y,7y}, another way to get the states with the given angular momentum and energy is by the Fourier transformation. The two ways are equivalent. In an arbitrary CFT that is not necessary free, for the operator $O(x)$ which is analytic at $x=0$
\begin{eqnarray}
	O(x) \nonumber&=&  \sum^{\infty}_{n=0} \frac{1}{n!}x^{\mu_{1}}\cdots x^{\mu_{n}}\partial_{\mu_{1}}\cdots \partial_{\mu_{n}}O(0)
 \\ &=&  \sum^{\infty}_{n=0} \sum_{l=n,n-2\cdots} \alpha_{n,l}\; Y^{\mu_{1}\cdots \mu_{l}}r^{n}[ \partial_{\mu_{1}}\cdots \partial_{\mu_{l}}(\partial^{2})^{\frac{n-l}{2}} -\textnormal{traces}]O(0). 
\end{eqnarray}
Taking $r=e^{it}$, we have
\begin{equation}
\int dt \;e^{-int} \; 	\int_{S^{2}} d\Omega \;Y_{\mu_{1}\cdots \mu_{l}}\;O(x) \propto[ \partial_{\mu_{1}}\cdots \partial_{\mu_{l}}(\partial^{2})^{\frac{n-l}{2}} -\textnormal{traces}]O(0). 
\end{equation}
$[ \partial_{\mu_{1}}\cdots \partial_{\mu_{l}}(\partial^{2})^{\frac{n-l}{2}} -\textnormal{traces}]O(0)$ is obtained by the Fourier transformation.

Finally, let us check the degrees of freedom in spin $s$ field. 
\begin{eqnarray}
	O_{i_{1}\cdots i_{s}}(x) &=&  \sum^{\infty}_{n=0} \frac{1}{n!}x^{\mu_{1}}\cdots x^{\mu_{n}}\partial_{\mu_{1}}\cdots \partial_{\mu_{n}}O_{i_{1}\cdots i_{s}}(0) . 
\end{eqnarray}
$O_{i_{1}\cdots i_{s}}(x)$ is conserved, $\partial^{i_{1}}O_{i_{1}\cdots i_{s}}(x)=0$, reducing the original $2s+1$ degrees of freedom to $2$. $O_{i_{1}\cdots i_{s}}(x)$ automatically obeys some kind of ``Coulomb gauge''. 
\begin{equation}
\partial_{i_{1}}	\partial_{\mu_{1}}\cdots \partial_{\mu_{n}}O_{i_{1}\cdots i_{s}}(0) = 0,
\end{equation}
so
\begin{equation}
(-i)^{n}[\partial_{\alpha} \partial_{\mu_{1}}\cdots \partial_{\mu_{l-1}}(\partial^{2})^{\frac{n-l}{2}} -\textnormal{traces}]O_{\alpha i_{1}\cdots i_{s-1}}(0) = O^{n}_{\alpha \mu_{1}\cdots \mu_{l-1};\;\alpha i_{1}\cdots i_{s-1}}(0) =0.
\end{equation} 
This is the manifestation of the gauge condition in momentum space. $O^{n}_{\mu_{1}\cdots \mu_{l};\;i_{1}\cdots i_{s}}(0)$, the creation operator for the particle with the quantum number $(n,l;s+1,s)$, is symmetric with
respect to the $\mu$ and $\nu$ indices separately and traceless with respect to all indices, i.e. 
\begin{equation}
	O^{n}_{\mu_{1}\cdots \mu_{l};\;i_{1}\cdots i_{s}}(0)=O^{n}_{\{\mu_{1}\cdots \mu_{l}\};\;i_{1}\cdots i_{s}}(0)=O^{n}_{\mu_{1}\cdots \mu_{l};\;\{i_{1}\cdots i_{s}\}}(0),
\end{equation}
\begin{equation}\label{147i}
	O^{n}_{\alpha \alpha \mu_{1}\cdots \mu_{l-2};\;i_{1}\cdots i_{s}}(0)=O^{n}_{\alpha \mu_{1}\cdots \mu_{l-1};\;\alpha i_{1}\cdots i_{s-1}}(0)=O^{n}_{\mu_{1}\cdots \mu_{l};\;\alpha \alpha i_{1}\cdots i_{s-2}}(0)=0. 
\end{equation}
$O^{n}_{\mu_{1}\cdots \mu_{l};\;i_{1}\cdots i_{s}}(0)$ only creates the physical states with no extra degrees of freedom to remove. There is no local gauge symmetry anymore. All we have is the global non-abelian higher spin symmetry. Also, 
\begin{equation}
\partial^{i_{1}}O'_{i_{1}\cdots i_{s}}(x)=\partial^{i_{1}} [U	O_{i_{1}\cdots i_{s}}(x)U^{+}]=U\partial^{i_{1}} 	O_{i_{1}\cdots i_{s}}(x)U^{+}=0,
\end{equation}
the local gauge condition is invariant under the global gauge transformation.

Since the higher spin charges can be explicitly constructed in the $3d$ CFT, it is direct to consider the action of the charges on the 1-particle state $\left|n,l;s+1, s \right\rangle$ and the higher spin transformation of $\left|n,l;s+1, s \right\rangle$. Before it, let us first see the realization of the singletons and the doubletons \cite{11q,12q,13q} in the $3d$ CFT. $a^{+} \left| 0 \right\rangle$ is the lowest energy state of the Dirac singleton $Rac = \mathcal{D}(1/2,0)$. 
\begin{equation}
	P_{\mu_{1}}\cdots P_{\mu_{n}}a^{+} \left| 0 \right\rangle  \propto a^{+}_{\mu_{1}\cdots \mu_{n}} \left| 0 \right\rangle. 
\end{equation}
$\mathcal{D}(1/2,0)$ is the space generated by $\{a^{+}_{\mu_{1}\cdots \mu_{n}} \left| 0 \right\rangle|n=0,1,\cdots\}$, i.e.
\begin{equation}
	\mathcal{D}(1/2,0)=H_{\{a^{+}_{\mu_{1}\cdots \mu_{n}} \left| 0 \right\rangle|n=0,1,\cdots\}}.
\end{equation}
Higher spin charges with the weight $p$ is the sum of the terms like $a^{+}_{\mu_{1}\cdots \mu_{n+p} i_{1}\cdots i_{l}  }a_{\nu_{1}\cdots \nu_{n} i_{1}\cdots i_{l} }$, so obviously, $\mathcal{D}(1/2,0)$ forms the irreducible representation of the higher spin algebra and also $so(3,2)$.

The tensor product of two scalar singletons is
\begin{equation}
	\mathcal{D}(1/2,0) \otimes \mathcal{D}(1/2,0) = H_{\{a^{+}_{\mu_{1}\cdots \mu_{m}}a^{+}_{\nu_{1}\cdots \nu_{n}} \left| 0 \right\rangle|m,n=0,1,\cdots\}},
\end{equation}
which is also the irreducible representation of the higher spin algebra but is reducible with respect to $so(3,2)$. The 1-particle Hilbert space of the higher spin gravity in $AdS_{4}$ is
\begin{eqnarray}\label{123w}
\nonumber 	\oplus_{s=0,2,\cdots}	\mathcal{D}(s+1,s)  &=& H_{	\{	O^{n}_{\mu_{1}\cdots \mu_{l};\;i_{1}\cdots i_{s}}(0) \left|0\right\rangle |s=0,2,\cdots;n=0,1,\cdots;l=n,n-2\cdots \}} \\  &=& 	 H_{	\{ \left|n,l;s+1,s\right\rangle |s=0,2,\cdots;n=0,1,\cdots;l=n,n-2\cdots \}}.
\end{eqnarray}
There will be
\begin{eqnarray}\label{153f}
\nonumber 	\mathcal{D}(1/2,0) \otimes \mathcal{D}(1/2,0)  &=& H_{\{a^{+}_{\mu_{1}\cdots \mu_{m}}a^{+}_{\nu_{1}\cdots \nu_{n}} \left| 0 \right\rangle|m,n=0,1,\cdots\}} \\\nonumber  &=& H_{	\{	O^{n}_{\mu_{1}\cdots \mu_{l};\;i_{1}\cdots i_{s}}(0) \left|0\right\rangle |s=0,2,\cdots;n=0,1,\cdots;l=n,n-2\cdots \}} \\  &=& 	\oplus_{s=0,2,\cdots}	\mathcal{D}(s+1,s). 
\end{eqnarray}
We will prove (\ref{153f}) in the generic $D$ dimension. For compactness, (\ref{123w}) is rewritten as
\begin{equation}
	\oplus_{s=0,2,\cdots}	\mathcal{D}(s+1,s)=H_{	\{ P_{\mu_{1}}\cdots P_{\mu_{n}}	O_{i_{1}\cdots i_{s}}(0) \left|0\right\rangle |s=0,2,\cdots;n=0,1,\cdots \}}.
\end{equation}
\begin{equation}
	P_{\mu_{1}}\cdots P_{\mu_{n}}	O_{i_{1}\cdots i_{s}}(0)\left|0\right\rangle = P_{\{\mu_{1}}\cdots P_{\mu_{n}\}}	O_{i_{1}\cdots i_{s}}(0)\left|0\right\rangle=P_{\mu_{1}}\cdots P_{\mu_{n}}	O_{\{i_{1}\cdots i_{s}\}}(0)\left|0\right\rangle,
\end{equation}
\begin{equation}
		P_{\alpha}P_{\mu_{2}}\cdots P_{\mu_{n}}	O_{\alpha i_{2}\cdots i_{s}}(0)\left|0\right\rangle = P_{\mu_{1}}\cdots P_{\mu_{n}}	O_{\alpha\alpha i_{3}\cdots i_{s}}(0)\left|0\right\rangle=0.  
\end{equation}
In dimension $D$, the tensor $T_{\mu_{1}\cdots \mu_{n}} = T_{\{\mu_{1}\cdots \mu_{n}\}}$ has the independent components $A_{n} =\frac{(n+D-1)!}{(D-1)!n!} $, the tensor $T_{\mu_{1}\cdots \mu_{n}} = T_{\{\mu_{1}\cdots \mu_{n}\}}$ with $T_{\alpha\alpha \mu_{3}\cdots \mu_{n}} = 0$ has the independent components $B_{n} = A_{n}-A_{n-2} = \frac{(2n+D-2)(n+D-3)!}{(D-2)!n!}$.

For the given energy $N+D-2$, there are
\begin{eqnarray}
\nonumber &&	a^{+}_{\mu_{1}\cdots \mu_{N}}a^{+}\left|0\right\rangle, \; a^{+}_{\mu_{1}\cdots \mu_{N-1}}a^{+}_{\mu_{N}}\left|0\right\rangle,\;\cdots,\;	a^{+}_{\mu_{1}\cdots \mu_{N/2}}a^{+}_{\mu_{N/2+1}\cdots \mu_{N}}\left|0\right\rangle , \;\textnormal{for even} \;N
\\   && a^{+}_{\mu_{1}\cdots \mu_{N}}a^{+}\left|0\right\rangle, \; a^{+}_{\mu_{1}\cdots \mu_{N-1}}a^{+}_{\mu_{N}}\left|0\right\rangle,\;\cdots,\;	a^{+}_{\mu_{1}\cdots \mu_{(N+1)/2}}a^{+}_{\mu_{(N+1)/2+1}\cdots \mu_{N}}\left|0\right\rangle, \;\textnormal{for odd} \;N 
\end{eqnarray}
in $H_{\{a^{+}_{\mu_{1}\cdots \mu_{m}}a^{+}_{\nu_{1}\cdots \nu_{n}} \left| 0 \right\rangle|m,n=0,1,\cdots\}}$, with the total number of the degrees of freedom 
\begin{eqnarray}\label{158ij}
\nonumber &&	B_{N}+B_{1}B_{N-1}+B_{2}B_{N-2} +\cdots+ B_{\frac{N}{2}-1}B_{\frac{N}{2}+1} +B_{\frac{N}{2}}(B_{\frac{N}{2}}+1)/2, \;\;\;\;\;\;\textnormal{for even} \;\;N
\\   && 	B_{N}+B_{1}B_{N-1}+B_{2}B_{N-2} +\cdots+ B_{\frac{N-1}{2}}B_{\frac{N+1}{2}}, \;\;\;\;\;\;\textnormal{for odd} \;\;N . 
\end{eqnarray}
On the other hand, in $H_{	\{ P_{\mu_{1}}\cdots P_{\mu_{n}}	O_{i_{1}\cdots i_{s}}(0) \left|0\right\rangle |s=0,2,\cdots;n=0,1,\cdots \}}$, 
the states with energy $N+D-2$ are
\begin{eqnarray}
\nonumber &&	O_{i_{1}\cdots i_{N}}(0)\left|0\right\rangle,\;\;  P_{\mu_{1}}P_{\mu_{2}}	O_{i_{1}\cdots i_{N-2}}(0)\left|0\right\rangle,\;\; \cdots,\;\;P_{\mu_{1}}\cdots P_{\mu_{N}}	O (0)\left|0\right\rangle\;\;\textnormal{for even} \;N 
\\ \nonumber  && P_{\mu_{1}} O_{i_{1}\cdots i_{N-1}}(0)\left|0\right\rangle,\;\;  P_{\mu_{1}}P_{\mu_{2}}P_{\mu_{3}}	O_{i_{1}\cdots i_{N-3}}(0)\left|0\right\rangle,\;\; \cdots, \;\; P_{\mu_{1}}\cdots P_{\mu_{N}}	O (0)\left|0\right\rangle\;\; \textnormal{for odd} \;N \\   &&  \;\;\;
\end{eqnarray}
with the total number of the degrees of freedom 
\begin{eqnarray}\label{158i}
\nonumber &&	B_{N}+(A_{2}B_{N-2}-A_{1}B_{N-3})+(A_{4}B_{N-4}-A_{3}B_{N-5})
+\cdots+ A_{N} , \;\;\;\;\;\;\textnormal{for even} \;\;N
\\\nonumber   && (A_{1}B_{N-1}-B_{N-2}) +(A_{3}B_{N-3}-A_{2}B_{N-4})\\   && +(A_{5}B_{N-5}-A_{4}B_{N-6}) +\cdots+ A_{N}, \;\;\;\;\;\;\textnormal{for odd} \;\;N .
\end{eqnarray}
(\ref{158ij}) and (\ref{158i}) are equal, so we arrive at (\ref{153f}). $H_{\{	\left|n,l;s+1,s\right\rangle |s=0,2,\cdots;n=0,1,\cdots;l=n,n-2\cdots \}}$, the 1-particle Hilbert space of the higher spin gravity in $AdS_{D+1}$, forms the irreducible representation of the higher spin algebra, but the reducible representation of $so(D,2)$, under which it is decomposed into the irreducible representations $\mathcal{D}(s+1,s)$:
\begin{equation}
	\oplus_{s=0,2,\cdots}	\mathcal{D}(s+1,s)=H_{	\{	\left|n,l;s+1,s\right\rangle |s=0,2,\cdots;n=0,1,\cdots;l=n,n-2\cdots \}},
\end{equation}
where
\begin{equation}
	\mathcal{D}(s+1,s) = H_{	\{	\left|n,l;s+1,s\right\rangle |n=0,1,\cdots;l=n,n-2\cdots \}}
\end{equation}
is the 1-particle Hilbert space for the spin $s$ field.

In \cite{12q,13q},
\begin{equation}
	\mathcal{D}(1/2,0) \otimes \mathcal{D}(1/2,0) = 	\oplus_{s=0,1,\cdots}	\mathcal{D}(s+1,s)
\end{equation}
was proved for $AdS_{4}$, while in \cite{16ty}, the discussion is extended to the arbitrary dimensions. In \cite{16ty}, the higher spin charges get the oscillator representation in terms of $a^{A}$, $\overline{a}^{A}$, satisfying the commutation relations
\begin{equation}
	[a^{A},\overline{a}^{B}]=-\eta^{AB},\;\;\;\;\;[a^{A},a^{B}]=0,\;\;\;\;\; [\overline{a}^{A},\overline{a}^{B}]=0,
\end{equation}
$(a^{A})^{+} = \overline{a}^{A}$, for $so(D,2)$, $A = 0,1,\cdots,D+1$. The single particle states of the higher spin gravity are then represented by the Fock module $\left|\Phi\right\rangle = \Phi(a^{A},\bar{a}^{A})\left|0\right\rangle$. Here, the higher spin charges and the single particle states are realized in terms of the oscillators $a^{+}_{\mu_{1}\cdots \mu_{n}}$ and $a_{\mu_{1}\cdots \mu_{n}}$ in CFT.

$H_{\{\left|n,l;s+1, s \right\rangle |s=0,2,\cdots;n=0,1,\cdots;l=n,n-2\cdots\}}$ forms the irreducible representation of the higher spin algebra. For the higher spin charge $Q$ with the dimension $\Delta$, there will be
\begin{equation}\label{e458}
Q	\left|n,l;s+1, s \right\rangle =\sum_{l', s'}Q^{n,l, s}_{l', s'}\left|s+n+\Delta-s',l';s'
+1, s' \right\rangle
\end{equation}
as well as
\begin{equation}
[Q,	O^{n}_{l;s} (0)] =\sum_{l', s'}Q^{n,l, s}_{l', s'}  O^{s+n+\Delta-s'}_{l';s'} (0). 
\end{equation}
Since $O_{s} (x) =e^{iP_{i}x^{i}}O^{0}_{0;s} (0) e^{-iP_{i}x^{i}}$, assuming 
\begin{equation}
	[[P_{i_{k}},\cdots[P_{i_{1}},Q]\cdots],O^{0}_{0;s} (0) ] = \sum_{l', s'}  q^{i_{1}\cdots i_{k}}_{l', s';s}    O^{k+s+\Delta-s'}_{l';s'} (0),
\end{equation}
\begin{eqnarray}
\nonumber && [Q,	O_{s} (x)] \\\nonumber &=& \sum_{l', s'}Q^{0,0, s}_{l', s'}  O^{s+\Delta-s'}_{l';s'} (x) \\\nonumber  && +\; e^{iP_{i}x^{i}}\{ -i x^{i}[[P_{i},Q],O^{0}_{0;s} (0)] +\frac{(-i)^{2}}{2!}x^{j}x^{i}[[P_{j},[P_{i},Q]],O^{0}_{0;s} (0) ]+\cdots   \}e^{-iP_{i}x^{i}} \\\nonumber &=& \sum_{l', s'}Q^{0,0, s}_{l', s'}  O^{s+\Delta-s'}_{l';s'} (x) \\   && -\; i x^{i} \sum_{l', s'}  q^{i}_{l', s';s}    O^{1+s+\Delta-s'}_{l';s'} (x) +\frac{(-i)^{2}}{2!}x^{j}x^{i}\sum_{l', s'}  q^{ij}_{l', s';s}    O^{2+s+\Delta-s'}_{l';s'} (x)+\cdots   . 
\end{eqnarray}
The descendants $O^{n}_{l;s} (x) \sim \partial^{n}O_{s} (x)$. $O_{s} (x)$ as well as its descendants also form the irreducible representation of the higher spin algebra.

$\{\left|n,l;s+1, s \right\rangle\}$ is the orthonormal basis in the 1-particle Hilbert space of the higher spin gravity. (\ref{e458}) then gives a matrix realization of the higher spin algebra $ho(1|2:[3,2])$. Especially, $so(3,2)$ are block diagonal matrices for the basis 
\begin{equation}
	\{\left|n,l;1, 0 \right\rangle,\left|n,l;3, 2 \right\rangle,\left|n,l;5, 4 \right\rangle , \cdots \}.
\end{equation}
For a conserved charge $Q$, $Q^{+}$ is also the conserved charge, which means if the matrix $t\in ho(1|2:[3,2])$, $t^{+}\in ho(1|2:[3,2])$. Since the group parameters of $G[ho(1|2:[3,2])]$ are all real, with $t\pm t^{+}$ taking the place of $t$ and $t^{+}$, we will get a real algebra with $T=T^{+}$.

Now we will consider the inner products between the higher spin gravity states in AdS. For the single particle state, the checking with the low spin states shows that for $s\neq s'$, 
\begin{equation}
\left\langle  n',l';s'+1, s'	| n,l;s+1, s \right\rangle  =  0.
\end{equation}
In fact, since the quadratic Casimir operator $C_{2}$ is Hermitian,  
\begin{eqnarray}\label{138}
\nonumber && \left\langle  n',l';s'+1, s'|C_{2}	|n,l;s+1, s \right\rangle\\\nonumber  &=& 2(s^{2}-1)\left\langle  n',l';s'+1, s'	|n,l;s+1, s \right\rangle\\  &=&2(s'^{2}-1)\left\langle  n',l';s'+1, s'|n,l;s+1, s \right\rangle . 
\end{eqnarray}
The 1-particle Hilbert spaces with different spins are orthogonal to each other. The generic $m$-particle states can be written as 
\begin{equation}
\frac{1}{N^{\frac{m}{2}}} O^{n_{1}}_{\mu^{1}_{1}\cdots \mu^{1}_{l_{1}};\;i^{1}_{1}\cdots i^{1}_{s_{1}}}(0)\cdots	O^{n_{m}}_{\mu^{m}_{1}\cdots \mu^{m}_{l_{m}};\;i^{m}_{1}\cdots i^{m}_{s_{m}}}(0) \left| 0 \right\rangle, 
\end{equation}
which has the energy $\sum^{m}_{i=1} (s_{i}+n_{i}+1)$. The normalization factor $1/N^{\frac{m}{2}}$ is added because the norm of $P^{(n)}_{\mu_{1}\cdots \mu_{l}} O^{++}_{i_{1}\cdots i_{s}}(0)\left| 0 \right\rangle$ is proportional to $N^{\frac{1}{2}}$. 
\begin{equation}
\frac{1}{N}O^{1}_{\mu}(0) 	O^{1}_{\nu}(0)  \left| 0 \right\rangle = -\frac{4}{3N}a_{\mu}^{+p}a^{+p}a_{\nu}^{+q}a^{+q}\left| 0 \right\rangle
\end{equation}
is the state for two spin $0$ particles, each has the angular momentum $1$. 
\begin{equation}
\frac{1}{N}O^{0}(0)  O^{0}_{i_{1}i_{2}}(0)  \left| 0 \right\rangle =\frac{1}{N}a^{+r}a^{+r} (\frac{2}{\sqrt{5}}a^{+s}a^{+s}_{i_{1}i_{2}}-a^{+s}_{i_{1}}a^{+s}_{i_{2}}+\frac{1}{3}\delta_{i_{1}i_{2}}a^{+s}_{\alpha}a^{+s}_{\alpha})\left|0\right\rangle
\end{equation}
is the state for one spin $0$ particle and one spin $2$ particle, both have the zero orbit angular momentum. $p$, $q$, $r$, $s$ are color indices.

We can also calculate the inner products between the multi-particle states. As we have seen before, states created by $ O^{n}_{\mu_{1}\cdots \mu_{l};\;i_{1}\cdots i_{s}}(0)$ automatically carry the right degrees of freedom, so the corresponding inner products give the physical amplitude. It is expected that with the higher spin gravity in $AdS_{4}$ quantized and the multi-particle eigenstates constructed, such inner products can be recovered.    
\begin{equation}\label{jo6}
\frac{1}{N^{2}}\left\langle 0\right| O^{+1}_{\mu}(0) 	O^{+1}_{\nu}(0) O^{0}(0)  O^{0}_{i_{1}i_{2}}(0) \left| 0 \right\rangle=\frac{12}{N}(3\delta_{i_{1}\rho}\delta_{i_{2}\sigma} + 3\delta_{i_{2}\rho}\delta_{i_{1}\sigma}- 2\delta_{i_{1}i_{2}}\delta_{\rho\sigma}     ), 
\end{equation}
so the 
\begin{equation}\label{142}
	\textnormal{spin}\; 0 + \textnormal{spin} \;0 \longleftrightarrow \textnormal{spin} \;0 + \textnormal{spin} \;2
\end{equation}
process is possible. Similarly, there is also
\begin{equation}\label{143}
	\textnormal{spin}\; 4 + \textnormal{spin} \;0 \longleftrightarrow \textnormal{spin} \;2 + \textnormal{spin} \;2.
\end{equation}
(\ref{jo6}) is proportional to $1/N$, while the cubic coupling constant in higher spin gravity is $g \sim 1/\sqrt{N}$, so (\ref{142}) is the combination of two cubic vertices. When $N \rightarrow \infty$, the gravity theory is free; the inner product (\ref{jo6}) becomes zero. This is expected, since at the free theory level, $O^{1}_{\mu}(0)\left| 0 \right\rangle 		O^{1}_{\nu}(0)  \left| 0 \right\rangle$ and $O^{0}(0)  \left| 0 \right\rangle O^{0}_{i_{1}i_{2}}(0)   \left| 0 \right\rangle$ are orthogonal to each other. Since the CFT is free, states with the different numbers of particles are orthogonal to each. However, the theory in AdS is not free and the spin changing interactions do exist.

Multi-particle states also form the representation of the higher spin algebra.
\begin{eqnarray}
\nonumber && Q  O^{n_{1}}_{\mu^{1}_{1}\cdots \mu^{1}_{l_{1}};\;i^{1}_{1}\cdots i^{1}_{s_{1}}}(0)\cdots	O^{n_{m}}_{\mu^{m}_{1}\cdots \mu^{m}_{l_{m}};\;i^{m}_{1}\cdots i^{m}_{s_{m}}}(0) \left| 0 \right\rangle \\\nonumber  & = & [Q ,O^{n_{1}}_{\mu^{1}_{1}\cdots \mu^{1}_{l_{1}};\;i^{1}_{1}\cdots i^{1}_{s_{1}}}(0)]\cdots	O^{n_{m}}_{\mu^{m}_{1}\cdots \mu^{m}_{l_{m}};\;i^{m}_{1}\cdots i^{m}_{s_{m}}}(0) \left| 0 \right\rangle \\\nonumber &&+ \; \cdots \\\nonumber &&+ \; O^{n_{1}}_{\mu^{1}_{1}\cdots \mu^{1}_{l_{1}};\;i^{1}_{1}\cdots i^{1}_{s_{1}}}(0)\cdots  [Q,	O^{n_{m}}_{\mu^{m}_{1}\cdots \mu^{m}_{l_{m}};\;i^{m}_{1}\cdots i^{m}_{s_{m}}}(0)  ]  \left| 0 \right\rangle. 
\end{eqnarray}
In this respect, they can be taken as the direct product of the single particle states, although they are not the direct products as we can see from the calculation of the inner products.

Finally, let us have a complete classification for the states in the Hilbert space of the $O(N)$ vector model. Imposing the condition that the physical states must be $O(N)$ invariant. The Fock space of the theory is 
\begin{equation}\label{ft5}
	\oplus^{\infty}_{k=0} H_{\{a^{+i_{1}}_{\mu^{1}_{1}\cdots \mu^{1}_{n_{1}}} a^{+i_{1}}_{\nu^{1}_{1}\cdots \nu^{1}_{m_{1}}}  \cdots a^{+i_{k}}_{\mu^{k}_{1}\cdots \mu^{k}_{n_{k}}} a^{+i_{k}}_{\nu^{k}_{1}\cdots \nu^{k}_{m_{k}}} \left| 0 \right\rangle |i_{p}=1,\cdots,N;\; m_{p},n_{p}=0,1,\cdots,\infty\}}.  
\end{equation}
With a change of the basis, (\ref{ft5}) becomes the Fock space of the higher spin gravity in $AdS_{4}$. Since the gauge condition is intrinsically followed, the states are all physical with no extra gauge degrees of freedom. In N=4 SYM theory, except for the trace, the $SU(N)$ invariant operators can also be constructed as the determinant \cite{14q1,14q}, which are dual to the giant gravitons (spherical branes) extending in $AdS_{5}$ or $S^{5}$ \cite{15q,17q,16q}. However, in $O(N)$ vector model, the determinant operators are $SO(N)$ other than $O(N)$ invariant, thus do not exist in the $O(N)$ invariant spectrum.

\section{Bulk operator from CFT revisited}

We have considered the CFT definition of the quantum higher spin gravity in $AdS_{4}$. For the generic AdS/CFT correspondence, the discussion is the same. Since the CFT and the AdS theory share the same symmetry, one can compute the conserved charges in CFT, which could be identified with the corresponding charges for the theory in AdS. With the $so(3,2)$ generators given, in the radial quantization of the CFT, we will first find a set of the primary operators $\{O(x)\}$ with $[K_{\mu},O(0)]=0$, $[D,O(0)]=-i\Delta_{O} O(0)$ as well as the descendants $O^{n}_{l}(0)$. $\{O^{n_{1}}_{l_{1}}(0) \cdots O^{n_{m}}_{l_{m}}(0)\left|0\right\rangle\}$ gives the Fock space for the physical particles in AdS theory. The inner products between the fock states encode the dynamical information. Constructing $O(0)$ as well as its derivatives at $0$ requires to solve the CFT exactly, which is not so straightforward in the interacting theory. Getting rid of these technical difficulties, the philosophy is identical. CFT could offer the spectrum as well as the physical particle states in AdS. A natural question is whether there is more AdS information, for example, the bulk operators, which could be extracted from CFT. The problem has been intensively discussed, see, for example, \cite{10y}-\cite{9qwa2}. For higher spin situation, see \cite{1qs,2qs}, in which the bulk higher spin fields are constructed in terms of the bi-local fields in the $O(N)$ CFT \cite{3qs}. We will add more comment on the topic. The following discussion applies for the generic CFT that is not necessarily free.

$AdS_{4}$ could be taken as the hyperboloid in $R^{3,2}$
\begin{equation}\label{ftgy}
\sum^{4}_{a,b=0}\eta_{ab}	y^{a}y^{b} = y^{2}_{0}-\sum^{3}_{i=1}y^{2}_{i}+ y^{2}_{4} = 1,
\end{equation}
where $\eta_{ab}=\textnormal{diag}(+,(-)^{3},+)$. In terms of the coordinate $y^{a}$, the Killing vectors in $AdS_{4}$ can be written as
\begin{equation}\label{89io}
	L_{ab} = y_{a}\frac{\partial}{\partial y^{b}}-y_{b}\frac{\partial}{\partial y^{a}}. 
\end{equation}
The field theory in $AdS_{4}$ has the $SO(3,2)$ generators $M_{ab}$, which, when acting on the operator $\Phi_{\theta,s}(y)$, gives
\begin{equation}
	[M_{ab}, \Phi_{\theta,s}(y)]=(-iL_{ab}+\Sigma_{ab})\Phi_{\theta,s}(y). 
\end{equation}
$\Phi_{\theta,s}(y)$ is the operator with the spin $s$. The meaning of $\theta$ will be explained later. Note that $M_{ab}$ is coordinate dependent: under the $SO(3,2)$ transformation, both $M_{ab}$ and $y$ will change. In the given coordinate system, consider a special point $o = (1,0,0,0,0)$, $i=1,2,3$, 
\begin{eqnarray}\label{4rt7}
\nonumber&& [M_{0i}, \Phi_{\theta,s}(o)]=-i \frac{\partial}{\partial y^{i}}\Phi_{\theta,s}(o), 
 \\\nonumber &&  [M_{04}, \Phi_{\theta,s}(o)]=-i \frac{\partial}{\partial y^{4}}\Phi_{\theta,s}(o), \\ && [M_{ij}, \Phi_{\theta,s}(o)]=[M_{i4}, \Phi_{\theta,s}(o)]=0,
\end{eqnarray}
where we have assumed $\Phi_{\theta,s}(o)$ only carries the spin indices in $(1,2,3)$ space. $\Sigma_{ij}$ transforms different components of $\Phi_{\theta,s}(o)$ into each other thus do not bring the substantial changes to the operator, so we simply let $[M_{ij}, \Phi_{\theta,s}(o)]=0$. $\Phi_{\theta,s}(o)$ commutes with the subalgebra $so(3,1)$ generated by $\{M_{ij},M_{i4} \}$. The action of $SO(3,2)$ in $AdS_{4}$ is transitive, so $\forall \; \Phi_{\theta,s}(y)$, $\exists \; g(y) \in SO(3,2)$, $\Phi_{\theta,s}(y) = g(y)\Phi_{\theta,s}(o)g(y)^{-1}$, (\ref{4rt7}) still holds with $\Phi_{\theta,s}(o)$ and $M_{ab}$ replaced by $\Phi_{\theta,s}(y)$ and $g(y)M_{ab}g(y)^{-1}$. 
\begin{equation}
	SO(3,2)/SO(3,1)\cong AdS_{4}.
\end{equation}
Starting from $\Phi_{\theta,s}(o)$ and $M_{ab}$,  
\begin{equation}\label{174r}
		\{\Phi_{\theta,s}(y)|y \in AdS_{4}\}=\{g \Phi_{\theta,s}(o) g^{-1}|g\in SO(3,2)\}
\end{equation}
gives the whole set of $\Phi_{\theta,s}(y)$ in $AdS_{4}$. The $EAdS_{4}$ is the hyperboloid in $R^{4,1}$, for which, one can just replace $y^{0}$ by $iy^{0}$ in the above discussion.

The boundary of $AdS_{4}$ could be characterized by $\sum^{3}_{i=1}y^{2}_{i} =R^{2}=\infty$, $y^{2}_{0}+ y^{2}_{4}= R^{2}+1$, which has the topology of $S^{1}\times S^{2}$. Consider a point $b \in \partial AdS_{4}$, $b=(\sqrt{R^{2}+1}, R,0,0,0)$. Let $\alpha, \beta = 2,3,4$, 
\begin{eqnarray}
\nonumber&& [M_{0\alpha}, \Phi_{\theta,s}(b)]=-i \sqrt{R^{2}+1} \frac{\partial}{\partial y^{\alpha}}\Phi_{\theta,s}(b)  , 
 \\\nonumber &&  [M_{1\alpha}, \Phi_{\theta,s}(b)]=i R \frac{\partial}{\partial y^{\alpha}}\Phi_{\theta,s}(b), 
\\\nonumber &&  [M_{01}, \Phi_{\theta,s}(b)]= -i (\sqrt{R^{2}+1}\frac{\partial}{\partial y^{1}}+R\frac{\partial}{\partial y^{0}})\Phi_{\theta,s}(b),
 \\ && [M_{\alpha \beta }, \Phi_{\theta,s}(b)]=0,
\end{eqnarray}
then for $R=\infty$, 
\begin{equation}\label{156t}
	K_{\alpha} = M_{0\alpha}+M_{1\alpha},\;\;\;\;\;\;\;\;\;P_{\alpha} = M_{0\alpha}-M_{1\alpha}, 
\end{equation}
\begin{equation}
	[K_{\alpha}, \Phi_{\theta,s}(b)]=0, \;\;\;[P_{\alpha}, \Phi_{\theta,s}(b)]=-2i R \frac{\partial}{\partial y^{\alpha}}\Phi_{\theta,s}(b), \;\;\;[M_{01}, \Phi_{\theta,s}(b)]= -i R(\frac{\partial}{\partial y^{0}}+\frac{\partial}{\partial y^{1}})\Phi_{\theta,s}(b). 
\end{equation}
$\Phi_{\theta,s}(b)$ commutes with the subalgebra $a[G_{3,1}]$ generated by $\{K_{\alpha},M_{\alpha \beta }\}$, which, however, is not isomorphic to $so(3,1)$.

$\Phi_{\theta,s}(b)$ and $\Phi_{\theta,s}(o)$, $\{K_{\alpha},M_{\alpha \beta }\}$ and $\{M_{ij},M_{i0} \}$ may be related by an infinite $SO(3,2)$ transformation $g(\infty)$. The finite $SO(3,2)$ transformation will map $\partial AdS_{4}$ to $\partial AdS_{4}$. Especially, the finite rotation generated by $M_{01}$ will leave $b$ invariant, so $[M_{01}, \Phi_{\theta,s}(b)]= \theta \Phi_{\theta,s}(b)$, $e^{iM_{01}\rho}\Phi_{\theta,s}(b)e^{-iM_{01}\rho} = e^{i\theta \rho}\Phi_{\theta,s}(b)$ only brings a change of the phase, which gives the origin of $\theta$. As a result, 
\begin{eqnarray}
\nonumber&&  \{g \Phi_{\theta,s}(b) g^{-1}|g\in SO(3,2)\}	 = \{e^{i(P_{\alpha}u^{\alpha}+M_{01}\rho)} \Phi_{\theta,s}(b) e^{-i(P_{\alpha}u^{\alpha}+M_{01}\rho)}\}
 \\ &=& \{e^{iP_{\alpha}x^{\alpha}} \Phi_{\theta,s}(b) e^{-iP_{\alpha}x^{\alpha}}|x \in \partial AdS_{4}\} = \{\Phi_{\theta,s}(x)|x \in \partial AdS_{4}\}, 
\end{eqnarray}
where we have used $e^{i(P_{\alpha}u^{\alpha}+M_{01}\rho)} = e^{iP_{\alpha}u^{\alpha}(1+\rho/2)}e^{iM_{01}\rho}$. Since
\begin{equation}
	\Phi_{\theta,s}(x)=e^{iP_{\alpha}x^{\alpha}} \Phi_{\theta,s}(b) e^{-iP_{\alpha}x^{\alpha}},
\end{equation}
$[P_{\alpha},\Phi_{\theta,s}(x) ]=-i	\partial_{\alpha}\Phi_{\theta,s}(x)$. Moreover, from the $so(3,2)$ algebra and the relation 
\begin{equation}\label{5rt645}
[K_{\alpha}, \Phi_{\theta,s}(b)]=[M_{\alpha \beta }, \Phi_{\theta,s}(b)]=0,\;\;\;	[M_{01}, \Phi_{\theta,s}(b)]=\theta \Phi_{\theta,s}(b),
\end{equation}
one can further get
\begin{eqnarray}\label{5rt6}
\nonumber&& [M_{01}, \Phi_{\theta,s}(x)]=	(\theta+x^{\alpha}\partial_{\alpha})\Phi_{\theta,s}(x), 
 \\\nonumber &&  [K_{\alpha}, \Phi_{\theta,s}(x)]=i[	2x_{\alpha}\theta-x^{2}\partial_{\alpha}+2 x_{\alpha}x^{\beta}\partial_{\beta}	] \Phi_{\theta,s}(x),  
 \\ && [M_{\alpha\beta}, \Phi_{\theta,s}(x)]=-i(x_{\alpha}\partial_{\beta}-x_{\beta}\partial_{\alpha}	) \Phi_{\theta,s}(x), 
\end{eqnarray}
which is the conformal transformation law for the primary operators in CFT. The transformation in (\ref{5rt6}) is different from (\ref{89io}) because we use the coordinate $x^{\alpha}$ other than $y^{a}$ at the boundary. The boundary is in the $(234)$ space with the signature $(-,-,+)$. $y^{4}$ is the time direction. 
\begin{equation}\label{192w}
	(z, x^{\alpha})=(\frac{1}{y^{0}+y^{1}}, \frac{y^{\alpha}}{y^{0}+y^{1}})
\end{equation}
parametrize $AdS_{4}$. When $z = 1/(\sqrt{R^{2}+1}+ R)\rightarrow 0$, $(0, x^{\alpha})$ gives the boundary.

Instead of $M_{0\alpha}$ and $M_{1\alpha}$, let us consider $M_{0i}$ and $M_{4i}$.  $P_{i} =M_{0i}+i M_{4i}$ and $K_{i} = -M_{0i}+i M_{4i}$ form the graded algebra with $[M_{04},P_{i}]=-P_{i}$ and $[M_{04},K_{i}]=K_{i}$. 
\begin{equation}
	[M_{4i}, \Phi_{\theta,s}(y)]=-i(y_{4}\frac{\partial}{\partial y^{i}}-y_{i}\frac{\partial}{\partial y^{4}})\Phi_{\theta,s}(y), \;\;\;\;\;\;\;\;\; [M_{0i}, \Phi_{\theta,s}(y)]=-i(y_{0}\frac{\partial}{\partial y^{i}}-y_{i}\frac{\partial}{\partial y^{0}})\Phi_{\theta,s}(y).
\end{equation}
Looking for $\Phi_{\theta,s}(b')$ with $[K_{i}, \Phi_{\theta,s}(b')]=0$, we find that $\Phi_{\theta,s}(b')$ must be the operator living at $b'=(iR,0,0,0,R)=\lim_{R \rightarrow\infty} (iR,0,0,0,\sqrt{R^{2}+1})$. We are actually talking about $EAdS_{4}$ in $R^{4,1}$ using the $R^{3,2}$ language. 
\begin{equation}
	[P_{i}, \Phi_{\theta,s}(b')]=2R\frac{\partial}{\partial y^{i}}\Phi_{\theta,s}(b'), \;\;\;\;\;\;\;\;\; [M_{04}, \Phi_{\theta,s}(b')]=\theta \Phi_{\theta,s}(b').
\end{equation}
$e^{iM_{04}\rho}$ will leave $b'$ fixed thus can only bring a phase to $\Phi_{\theta,s}(b')$. $\Phi_{\theta,s}(b')$ commutes with the subalgebra generated by $\{M_{ij},K_{i}\}$. At $\partial EAdS_{4}$, $\Phi_{\theta,s}(x) = e^{iP_{i} x^{i}}\Phi_{\theta,s}(b')e^{-iP_{i} x^{i}}$ although $P_{i}$ is not the unitary operator now. Wick-rotating $y_{0}$, the embedding of $EAdS_{4}$ is 
\begin{equation}
\sum^{4}_{a,b=0}\eta_{ab}	y^{a}y^{b} = -y^{2}_{0}-\sum^{3}_{i=1}y^{2}_{i}+ y^{2}_{4} = 1.
\end{equation}
$EAdS_{4}$ is parametrized by 
\begin{equation}\label{196t}
	(z, x^{i})=(\frac{1}{y^{0}+y^{4}}, \frac{y^{i}}{y^{0}+y^{4}}).
\end{equation}
$z = 1/(\sqrt{R^{2}+1}+ R)\rightarrow 0$ gives the boundary $(0, x^{i})$, which is in the $(123)$ space with the signature $(-,-,-)$. $M_{04}$ is the global time generator in $EAdS_{4}$. According to (\ref{156t}), $M_{04} = (P_{4}+K_{4})/2$.

From a given bulk operator $\Phi_{\theta,s}(o)$ with $o \in \textnormal{int}(AdS_{4})$, (\ref{174r}) gives the whole set of operators in AdS. Unfortunately, with the boundary operator $\Phi_{\theta,s}(b)$ given, its bulk extension is not unique. The finite $SO(3,2)$ transformation can only send the boundary operators to the boundary. 
\begin{eqnarray}
 \nonumber	\Phi_{\theta,s}(x) &=& e^{iP^{\alpha} x_{\alpha}}\Phi_{\theta,s}(b)e^{-iP^{\alpha} x_{\alpha}} 
 \\ \nonumber &=& \Phi_{\theta,s}(b)+ ix_{\alpha_{1}}[P^{\alpha_{1}},\Phi_{\theta,s}(b)]-\frac{1}{2}x_{\alpha_{1}}x_{\alpha_{2}}[P^{\alpha_{1}},[P^{\alpha_{2}},\Phi_{\theta,s}(b)]]+\cdots  \\&=& A_{\theta,s}+ ix_{\alpha_{1}}A^{\alpha_{1}}_{\theta,s} -\frac{1}{2}x_{\alpha_{1}}x_{\alpha_{2}}A^{\alpha_{1}\alpha_{2}}_{\theta,s}+\cdots . 
\end{eqnarray}
$\{A_{\theta,s}, A^{\alpha_{1}}_{\theta,s},A^{\alpha_{1}\alpha_{2}}_{\theta,s}\cdots \}$ forms an irreducible representation of $so(3,2)$ with $A_{\theta,s}$ the lowest energy operator, i.e. $[K_{\mu},A_{\theta,s} ]=0$, $[D, A_{\theta,s}]=-i\theta A_{\theta,s}$. From (\ref{5rt645}), we have\footnote{$[M_{\alpha \beta }, \Phi_{\theta,s}(b)]=0$ is replaced by the more accurate $[M_{\alpha \beta }, \Phi_{\theta,s}(b)]=\Sigma_{\alpha \beta } \Phi_{\theta,s}(b)$.} 
\begin{equation}\label{se45}
	[C_{2},\Phi_{\theta,s}(b) ]\equiv \frac{1}{2}[M^{ab},[M_{ab},\Phi_{\theta,s}(b) ]] =[\theta(\theta-3)+s(s+1)]\Phi_{\theta,s}(b),
\end{equation}
$[g,C_{2}]=0$, $\forall g \in SO(3,2)$, so 
\begin{equation}
		[C_{2}, A^{\alpha_{1}\cdots \alpha_{k}}_{\theta,s}]=[\theta(\theta-3)+s(s+1)]A^{\alpha_{1}\cdots \alpha_{k}}_{\theta,s}.
\end{equation}
As one possible bulk extension, we take
\begin{equation}\label{ft67b}
	\Phi^{f}_{\theta,s}(y) = \sum_{k}  \varphi_{\alpha_{1}\cdots \alpha_{k}}(y)    A^{\alpha_{1}\cdots \alpha_{k}}_{\theta,s}~.~
\end{equation}
Requiring    
\begin{eqnarray}\label{y78}
 \nonumber&&	[M_{ab}, \Phi^{f}_{\theta,s}(y) ]=\sum_{k}  \varphi_{\alpha_{1}\cdots \alpha_{k}}(y) [M_{ab},   A^{\alpha_{1}\cdots \alpha_{k}}_{\theta,s}]
 \\ &=&  -i \sum_{k} L_{ab}  \varphi_{\alpha_{1}\cdots \alpha_{k}}(y)    A^{\alpha_{1}\cdots \alpha_{k}}_{\theta,s}=-iL_{ab}\Phi^{f}_{\theta,s}(y)
\end{eqnarray}
will fix $\Phi^{f}_{\theta,s}(y)$. The boundary $\Phi^{f}_{\theta,s}(y)$ is not arbitrary but should also satisfy the constraints (\ref{y78}), which, at $\partial AdS_{4}$, reduces to (\ref{5rt6}). $\Phi^{f}_{\theta,s}(y)\longrightarrow \Phi_{\theta,s}(x)$ at $\partial AdS_{4}$. Due to (\ref{ft67b}), 
\begin{equation}
	[C_{2}, \Phi^{f}_{\theta,s}(y)]=[\lambda(\lambda-3)+s(s+1)]\Phi^{f}_{\theta,s}(y) =\frac{1}{2}L_{ab}L^{ab}\Phi^{f}_{\theta,s}(y) =\square  \Phi^{f}_{\theta,s}(y)  .  
\end{equation}
$\square$ is the invariant d'Alambertian in $AdS_{4}$. $\Phi^{f}_{\theta,s}(y)$ satisfies the free field equation. Just as (\ref{174r}), we also have  
\begin{equation}
		\{\Phi^{f}_{\theta,s}(y)|y \in AdS_{4}\}=\{g \Phi^{f}_{\theta,s}(o) g^{-1}|g\in SO(3,2)\}
\end{equation}
for some bulk $\Phi^{f}_{\theta,s}(o)$. So for any local operator $\Phi_{\theta,s}(y) $ in AdS, from the boundary $\Phi_{\theta,s}(x) $, one can always find a consistent bulk extension $\Phi^{f}_{\theta,s}(y) $ by solving the free field equation. $\Phi^{f}_{\theta,s}(y)$ and $\Phi_{\theta,s}(y) $ are identical at the boundary but are different in the bulk.

In fact, let $\Phi(y)$ be the arbitrary local operator in $AdS_{4}$, which can always be written as
\begin{equation}
		\{\Phi (y)|y \in AdS_{4}\}=\{g \Phi (o) g^{-1}|g\in SO(3,2)\}.
\end{equation}
Then if $[C_{2},\Phi (o)]=m^{2}\Phi (o)$, since $[g,C_{2}]=0$, there will be 
\begin{equation}
	[C_{2},\Phi (y)]=m^{2}\Phi (y) = \square \Phi (y). 
\end{equation}
$\Phi (y)$ satisfies the free field equation with the mass $m$. For $\Phi^{f}_{\theta,s}(y)$, $[C_{2},\Phi^{f}_{\theta,s}(o)]=[\lambda(\lambda-3)+s(s+1)] \Phi^{f}_{\theta,s}(o)$; for the generic $\Phi_{\theta,s}(y)$, we do not have the bulk $\Phi_{\theta,s}(o)$ which is the eigenstate of $C_{2}$.

For the theory in AdS, we are interested in the Heisenberg operators $\Psi_{\theta,s}(x,\rho)$ for fundamental fields of the theory. At $\partial AdS_{4}$, $\Psi_{\theta,s}(x,\rho)$ reduces to $\Psi_{\theta,s}(x)$ with the $SO(3,2)$ transformation law given by (\ref{5rt6}). One can get $\Psi^{f}_{\theta,s}(x,\rho)$ from the boundary $\Psi_{\theta,s}(x)$, but obviously, $\Psi_{\theta,s}(x,\rho)$ does not satisfy the free field equation unless the theory is free. Except for $\Psi_{\theta,s}(x,\rho)$, another set of operators of interest are $\Phi^{+}_{\lambda,s}(x,\rho)$, which are operators creating the physical particles of the type $(\lambda,s)$ in AdS.

Consider the QFT in AdS. Based on the nonperturbative analysis, it has a unique vacuum $\left| \Omega \right\rangle$, the operators $H$, $P_{\mu}$, $K_{\nu}$, and $M_{\mu\nu}$ generating the $SO(3,2)$ transformation, the discrete single particle states $\left| \omega,j;\lambda,s \right\rangle$ as well as the multi-particle states $\left| \Omega,J;\Lambda,S \right\rangle$, which are eigenstates of $H$ and $M_{\mu\nu}$\footnote{Different from the previous notation, here, $\omega$/$\Omega$ and $j$/$J$ are total energy and the total angular momentum.}. $\left| \omega,j;\lambda,s \right\rangle$/$\left| \Omega,J;\Lambda,S \right\rangle$ are states for physical particles whose masses taking the measured values.   
\begin{equation}\label{890}
	1=\left| \Omega \right\rangle \left\langle \Omega\right|+ \sum_{\omega,j,\lambda,s}\left|\omega,j;\lambda,s \right\rangle\left\langle \omega,j;\lambda,s\right|+\sum_{\Omega,J,\Lambda,S}\left|\Omega,J;\Lambda,S \right\rangle\left\langle \Omega,J;\Lambda,S\right|. 
\end{equation}
Under the $SO(3,2)$ transformation, the Hilbert space is decomposed into the direct sum of the irreducible representations characterized by $(\lambda,s)$/$(\Lambda,S)$. In each representation, there is a lowest energy state $\left| \lambda,s;\lambda,s \right\rangle$/$\left| \Lambda,S;\Lambda,S \right\rangle$ with $K_{\mu}\left|  \lambda,s;\lambda,s \right\rangle = 0$/$K_{\mu} \left| \Lambda,S;\Lambda,S \right\rangle=0$.

It is possible to construct the creation operator $a^{+}_{ \omega,j;\lambda,s}$ with $a^{+}_{ \omega,j;\lambda,s} \left| \Omega \right\rangle= \left|  \omega,j;\lambda,s \right\rangle$, $a_{ \omega,j;\lambda,s} \left| \Omega \right\rangle= 0$. Operators creating the physical particles in AdS should take the form of 
\begin{equation}
	\Phi_{\lambda,s}^{+}(x,\rho) = \sum_{ \omega,j}\;\varphi_{ \omega,j;\lambda,s}(x,\rho)a^{+}_{ \omega,j;\lambda,s}.
\end{equation}
Again, $[M_{ab},\Phi_{\lambda,s}^{+}(x,\rho)]=-iL_{ab} \Phi_{\lambda,s}^{+}(x,\rho)$ will fix $\Phi_{\lambda,s}^{+}(x,\rho)$. 
\begin{equation}
	[C_{2},a^{+}_{ \omega,j;\lambda,s}]=[\lambda(\lambda-3)+s(s+1)] a^{+}_{ \omega,j;\lambda,s},
\end{equation}
so 
\begin{equation}
[C_{2},	\Phi_{\lambda,s}^{+}(x,\rho) ]=	\square \Phi_{\lambda,s}^{+}(x,\rho)=[\lambda(\lambda-3)+s(s+1)]\Phi_{\lambda,s}^{+}(x,\rho). 
\end{equation}
$\Phi_{\lambda,s}^{+}(x,\rho)$ satisfies the free field equation in $AdS$, which is entirely due to the symmetry. For interacting QFT in Minkowski spacetime, local operators creating the physical particle with mass $m$ should also satisfy the free field equation with mass $m$. $\Phi_{\lambda,s}^{+}(x,\rho)$ is a totally symmetric traceless tensor in $AdS_{4}$. The redundant degrees of freedom must be removed. For massless particles, one can impose the ``Coulomb gauge'' at $\rho = \pi/2$ and then solve the whole $\Phi_{\lambda,s}^{+}(x,\rho)$ from the equation of motion.

$\left| x,\rho \right\rangle_{\lambda,s}=\Phi_{\lambda,s}^{+}(x,\rho)\left| \Omega \right\rangle$ is the state for a spin $s$ particle at $(x,\rho)$ with the physical mass $[\lambda(\lambda-3)+s(s+1)]^{1/2}$, which does not need to be the bare mass in Lagrangian. $\Phi_{\lambda,s}^{+}(x,\rho)\left| \Omega \right\rangle$ should be distinguished from $\Psi_{\theta,s}(x,\rho)\left| \Omega \right\rangle$.   
\begin{eqnarray}
\Psi_{\theta,s}(x,\rho)\left| \Omega \right\rangle \nonumber &=&\left| \Omega \right\rangle \left\langle \Omega\right|\Psi_{\theta,s}(x,\rho)\left| \Omega \right\rangle+ \sum_{\omega,j,\lambda,s}\left|\omega,j;\lambda,s \right\rangle\left\langle \omega,j;\lambda,s\right|\Psi_{\theta,s}(x,\rho)\left| \Omega \right\rangle  \\ &+& \sum_{\Omega,J,\Lambda,S}\left|\Omega,J;\Lambda,S \right\rangle\left\langle \Omega,J;\Lambda,S\right|\Psi_{\theta,s}(x,\rho)\left| \Omega \right\rangle. 
\end{eqnarray}
$\Psi_{\theta,s}(x,\rho)$ satisfies the nonlinear equation. As a result, except for the single particle states in $(\lambda,s)$ representation, the rest states in (\ref{890}) may also contribute. Usually, $\Phi_{\lambda,s}^{+}(x,\rho)\left| \Omega \right\rangle\neq \Psi_{\theta,s}(x,\rho)\left| \Omega \right\rangle$.

One can also construct the multi-particle states $a^{+}_{ \omega_{1},j_{1};\lambda_{1},s_{1}}\cdots a^{+}_{ \omega_{k},j_{k};\lambda_{k},s_{k}}\left| \Omega \right\rangle$ and compute the inner products between them. If the commutation relation for $a^{+}$ and $a$ is just $[a_{ \omega,j;\lambda,s},a^{+}_{ \omega',j';\lambda',s'}]=i \delta_{\omega,\omega'}\delta_{j,j'}\delta_{\lambda,\lambda'}\delta_{s,s'}$, multi-particle states reduce to the products of the single particle states with the trivial inner products. This is not the case. $a^{+}_{ \omega,j;\lambda,s}$ and $a_{ \omega,j;\lambda,s}$, as the operators creating and annihilating the physical particles, are complicated composite operators of the bare fields and thus will have the nontrivial commutator. As a result, for spacelike separations, 
\begin{equation}
	[\Psi_{\theta,s}(x,\rho),\Psi_{\theta,s}(x',\rho')]=0
\end{equation}
but
\begin{equation}
	[\Phi_{\lambda,s}(x,\rho),\Phi_{\lambda,s}(x',\rho')]\neq 0
\end{equation}
unless the theory is free. $\Phi_{\lambda,s}(x,\rho) = \Phi^{-}_{\lambda,s}(x,\rho)+\Phi^{+}_{\lambda,s}(x,\rho)$. In position space,
\begin{equation}
	\left\langle \Omega\right|\Phi^{-}_{\lambda',s'}(x',\rho')\Phi_{\lambda,s}^{+}(x,\rho)\left| \Omega \right\rangle = \delta_{\lambda,\lambda'}\delta_{s,s'}\Delta (x',\rho';x,\rho)
\end{equation}
gives the transition amplitude for a physical particle $(\lambda,s)$ at $(x,\rho)$ to jump to the physical particle $(\lambda',s')$ at $(x',\rho')$, which is the same as the free field calculation. 
\begin{equation}\label{hui}
	\left\langle \Omega\right|\Phi^{-}_{\lambda'_{1},s'_{1}}(x'_{1},\rho'_{1})\cdots \Phi^{-}_{\lambda'_{m},s'_{m}}(x'_{m},\rho'_{m}) \Phi_{\lambda_{1},s_{1}}^{+}(x_{1},\rho_{1})\cdots \Phi_{\lambda_{n},s_{n}}^{+}(x_{n},\rho_{n})\left| \Omega \right\rangle 
\end{equation}
is the transition amplitude for $n$ physical particles $(\lambda_{1},s_{1})$, $\cdots$, $(\lambda_{n},s_{n})$ living at $(x_{1},\rho_{1})$, $\cdots$, $(x_{n},\rho_{n})$, to jump to $m$ physical particles $(\lambda'_{1},s'_{1})$, $\cdots$, $(\lambda'_{m},s'_{m})$ living at $(x'_{1},\rho'_{1})$, $\cdots$, $(x'_{m},\rho'_{m})$. We may assume $t_{1}=\cdots =t_{n}$, $t'_{1}=\cdots =t'_{m}$ so that (\ref{hui}) gives the time transition. (\ref{hui}) is not equal to the free theory value anymore. Multi-particles will interact with each other.

Back to CFT, the conformal symmetry is generated by $\{D,K_{\mu},P_{\mu},M_{\mu\nu}\}$. There is a complete set of primary operators $\{O\}$ with $[K_{\mu},O(0)]=0$. The operators can be further rearranged into the common eigenstates of $D$ and $M_{\mu\nu}M^{\mu\nu}/2$, i.e., $[D, O_{\lambda,s}]=-i \lambda O_{\lambda,s}$, $[M_{\mu\nu}M^{\mu\nu}/2, O_{\lambda,s}]=s(s+1) O_{\lambda,s}$, where $ O_{\lambda,s}$ can also be the linear combination of the single and multi-trace operators. We are interested in the operators with the definite conformal dimension, because only these operators could be taken as the boundary limit of the bulk operators in AdS.

For the boundary of $AdS_{4}$, we have discussed two situations, the Minkowski $(234)$ space and the Euclidean $(123)$ space. First, let us consider the single trace primary operator $O_{\lambda,s}(x)$ in Minkowski CFT. The conformal transformation law for $O_{\lambda,s}(x)$ is the same as the $SO(3,2)$ transformation law of $\Phi_{\theta,s}(x)$ at $\partial AdS$ with $\lambda=\theta$, so if $\{D,K_{\mu},P_{\mu},M_{\mu\nu}\}$ is identified as the $so(3,2)$ generators in AdS, $O_{\lambda,s}(x)$ could be taken as the boundary limit of some AdS operators. $\Phi_{\lambda,s}(x) = O_{\lambda,s}(x)$, where $x$ is in $(234)$ space as is parameterized in (\ref{192w}). It is possible to find a free theory bulk extension $O^{f}_{\lambda,s}(x,\rho)$ by solving the free field equation 
\begin{equation}
  \square  O^{f}_{\lambda,s}(x,\rho) =[\lambda(\lambda-3)+s(s+1)]O^{f}_{\lambda,s}(x,\rho).  
\end{equation}
However, from the given boundary operator, the bulk extension is not unique. We are interested in $\Psi_{\theta,s}(x,\rho)$. Even if $O_{\lambda,s}(x)$ could be taken as the boundary limit of $\Psi_{\theta,s}(x,\rho)$, there is no clue to find the bulk extension which is of course not $O^{f}_{\lambda,s}(x,\rho)$. In principle, as long as one bulk $\Psi_{\theta,s}(o)$ get the CFT realization, (\ref{174r}) will give the whole set of $\Psi_{\theta,s}(x,\rho)$ in AdS, since the $so(3,2)$ generators are constructed. Again, we do not know how to write $\Psi_{\theta,s}(o)$ in CFT.

Now, let us turn to $EAdS_{4}$ whose boundary is given by (\ref{196t}). It is the Euclidean CFT that should be put on the boundary. $0 \equiv \lim_{R\rightarrow \infty}(iR,0,0,0,\sqrt{R^{2}+1})$. $O^{+}_{\lambda,s}(0)$, the analytic part of $O_{\lambda,s}(0)$, or in the language of the radial quantization of the Minkowski CFT, the positive frequency part, can be identified as $a^{+}_{\lambda,s;\lambda,s}$. The descendants of $O^{+}_{\lambda,s}(0)$ then gives $a^{+}_{ \omega,j;\lambda,s}$.
\begin{equation}
	O^{+}_{\lambda,s}(x)=e^{iP_{\mu}x^{\mu}}O^{+}_{\lambda,s}(0)e^{-iP_{\mu}x^{\mu}} =e^{iP_{\mu}x^{\mu}}a^{+}_{\lambda,s;\lambda,s}e^{-iP_{\mu}x^{\mu}} 
\end{equation}
are boundary operators. The bulk extension $O^{+}_{\lambda,s}(x,\rho)$ can be obtained by solving the free field equation. $O^{+}_{\lambda,s}(x,\rho)$ are the operators creating physical particles in the bulk. $O^{+}_{\lambda,s}(x,\rho) = \Phi_{\lambda,s}^{+}(x,\rho)$. The CFT construction could offer $\Phi_{\lambda,s}^{+}(x,\rho)$. As we have seen before, for massless particles, $\partial O^{+}_{\lambda,s}(x)=0$, $O^{+}_{\lambda,s}(x,\rho)$ obeys the ``Coulomb gauge'' at the boundary and so the whole set of $O^{+}_{\lambda,s}(x,\rho)$ in the bulk has no extra degrees of freedom.

Multi-particle states $O_{n_{1},l_{1};\lambda_{1},s_{1}}^{+}(0)\cdots O_{n_{k},l_{k};\lambda_{k},s_{k}}^{+}(0) \left|0\right\rangle$ can also be constructed and are supposed to produce the same inner products as that of $a^{+}_{ \omega_{1},j_{1};\lambda_{1},s_{1}}\cdots a^{+}_{ \omega_{k},j_{k};\lambda_{k},s_{k}}\left| \Omega \right\rangle$. As we can check in $3d$ free CFT, the commutator $[O_{n,l;\lambda,s}^{+}(0),O_{n',l';\lambda',s'}^{-}(0)]$ is indeed nontrivial, for example, 
\begin{equation}
[O_{0,0;1,0}^{+}(0),O_{1,1;1,0}^{-}(0)]=[\frac{a^{+}a^{+}}{N^{1/2}},\frac{a_{\mu}a}{N^{1/2}}]=-\frac{2}{N}a^{+}a_{\mu}.
\end{equation}
It is only in $N\rightarrow \infty$ limit
\begin{equation}
	[O_{n,l;\lambda,s}^{+}(0),O_{n',l';\lambda',s'}^{-}(0)]\rightarrow  \delta_{n,n'}\delta_{l,l'}\delta_{\lambda,\lambda'}\delta_{s,s'}.
\end{equation}
The CFT realization of (\ref{hui}) is
\begin{equation}
	\left\langle 0\right|O^{-}_{\lambda'_{1},s'_{1}}(x'_{1},\rho'_{1})\cdots O^{-}_{\lambda'_{m},s'_{m}}(x'_{m},\rho'_{m}) O_{\lambda_{1},s_{1}}^{+}(x_{1},\rho_{1})\cdots O_{\lambda_{n},s_{n}}^{+}(x_{n},\rho_{n})\left| 0\right\rangle.  
\end{equation}
Again, assuming $t_{1}=\cdots =t_{n}$ and $t'_{1}=\cdots =t'_{m}$ gives the time transition. As we can see explicitly, $O^{+}_{\lambda,s}$ are composite operators, and so the multi-particle transition amplitude is nontrivial with the interaction playing a role.

It is commonly recognized that from CFT, the bulk operator $O_{\lambda,s}(x,\rho)=O^{+}_{\lambda,s}(x,\rho)+O^{-}_{\lambda,s}(x,\rho)$ can be built satisfying the free field equation. For spacelike $(x,\rho)$ and $(x',\rho')$, $[O_{\lambda,s}(x,\rho),O_{\lambda,s}(x',\rho')]=0$ when $N = \infty$, in which case, the locality can be preserved. When $N=\infty$, the bulk theory is free, $O_{\lambda,s}(x,\rho)$ can be identified as $\Psi_{\lambda,s}(x,\rho)$, while for the latter, the spacelike commuting condition is always imposed. However, the free theory is not interesting, we still want to consider the finite $N$ case, for which, $O_{\lambda,s}(x,\rho)\neq \Psi_{\lambda,s}(x,\rho)$, $[O_{\lambda,s}(x,\rho),O_{\lambda,s}(x',\rho')]\neq 0$. In \cite{qwert2,8qwa,qwert1}, it was proposed that the multi-trace operators can be added to $O_{\lambda,s}(x,\rho)$ so that the spacelike commuting condition is obeyed also for finite $N$, making the final operator a candidate of $\Psi_{\lambda,s}(x,\rho)$. It is good to get the CFT realization of $\Psi_{\lambda,s}(x,\rho)$. However, $O^{+}_{\lambda,s}(x,\rho)$ and $O^{-}_{\lambda,s}(x,\rho)$ also have their own physical significance: they are the operators creating and annihilating physical particles thus are directly related with the measurement. Since $[O^{-}_{\lambda,s}(x,\rho),O^{-}_{\lambda,s}(x',\rho')]=[O^{+}_{\lambda,s}(x,\rho),O^{+}_{\lambda,s}(x',\rho')]=0$, from $O^{-}_{\lambda,s}(x,\rho)$ one can also define the coherent states: 
\begin{equation}
	O^{-}_{\lambda,s}(x,\rho)\left|\alpha\right\rangle=f^{\alpha}_{\lambda,s}(x,\rho)\left|\alpha\right\rangle. 
\end{equation}
$[\square-\lambda(\lambda-3)-s(s+1)]O^{-}_{\lambda,s}=0$, so $[\square-\lambda(\lambda-3)-s(s+1)]f^{\alpha}_{\lambda,s}(x,\rho)=0$, $f^{\alpha}_{\lambda,s}(x,\rho)$ also satisfies the free field equation. In most cases, the masses in Lagrangian are very different from the physical masses. Consider a scalar field with the bare mass $m_{b}$ and the physical mass $m_{p}$. The observed classical scalar field always has the physical mass $m_{p}$ in the classical field equation, so it is necessary to construct the physical operators like $O^{+}_{\lambda,s}(x,\rho)$ to relate the underlying theory to the observation.

In some sense, the CFT definition of the quantum gravity is in the S-matrix formalism, although there is no S-matrix here. The spectrum, states and their inner products can all be constructed, but the dynamics of the fundamental quantum fields giving rise to these excitations is obscure. Equivalent to building $\Psi_{\lambda,s}(x,\rho)$, one can instead try to find 
\begin{equation}\label{g6}
	\left| f_{\lambda,s} \right\rangle= f_{\Omega}\left| \Omega \right\rangle +\sum_{\omega,j,\lambda,s}\;f_{\omega,j,\lambda,s}\left|\omega,j;\lambda,s \right\rangle+\sum_{\Omega,J,\Lambda,S}\;f_{\Omega,J,\Lambda,S}\left|\Omega,J;\Lambda,S \right\rangle,
\end{equation}
where $\left| f_{\lambda,s} \right\rangle$ are eigenstates of the field operators $\Psi_{\lambda,s}(x,\rho)$ with
\begin{equation}
	 \Psi_{\lambda,s}(x,\rho)	\left| f_{\lambda,s} \right\rangle=f_{\lambda,s}(x,\rho)	\left| f_{\lambda,s} \right\rangle 
\end{equation}
at $t=t_{0}$. Since $H$ is known in CFT, $\left\langle f_{\lambda',s'}'|e^{iH(t-t_{0})}|f_{\lambda,s}\right\rangle$ gives the transition amplitude. (\ref{g6}) is just a change of the basis from the energy eigenstates to the field value eigenstates. The coefficients in (\ref{g6}) can be easily determined only in the free AdS theory.

For higher spin gravity, it is possible to explicitly write the bulk operators creating physical particles from the CFT dual. Let us consider the simplest situation for $O^{+}_{1,0}$. For $o = (1,0,0,0,0)$, 
\begin{equation}\label{90ok}
	[M_{\mu\nu}, O^{+}_{1,0}(o)]=[M_{\mu 4}, O^{+}_{1,0}(o)]=[K_{\mu }-P_{\mu }, O^{+}_{1,0}(o)]=0.  
\end{equation}
$\{M_{04}, M_{0\mu}\} \subset K_{1,0}(o)$ generates the tangent space along $AdS_{4}$. From (\ref{90ok}), $O^{+}_{1,0}(o)$ is solved as 
\begin{equation}
O^{+}_{1,0}(o) = \sum \frac{m!}{2(2m+1)!!}	 a^{+}_{i_{1}\cdots i_{m}}a^{+}_{i_{1}\cdots i_{m}}. 
\end{equation}
$O^{+}_{1,0}(o)$ has several interesting properties.
\begin{equation}
\frac{1}{2}	[O^{-}_{1,0}(o), O^{+}_{1,0}(o)]= \hat{N} = \sum \frac{m!}{2(2m+1)!!}	 a^{+}_{i_{1}\cdots i_{m}}a_{i_{1}\cdots i_{m}} + \textnormal{const} . 
\end{equation}
Let $a^{+}$ and $a$ collectively represent $a_{i_{1}\cdots i_{l}}^{+}$ and $a_{j_{1}\cdots j_{l'}}$ respectively, 
\begin{equation}
	[\hat{N}, (a^{+})^{m}a^{n}]=\frac{1}{2}(m-n) (a^{+})^{m}a^{n},
\end{equation}
\begin{equation}
	(\hat{N}-\textnormal{const})(a^{+}a^{+})^{n} \left|0\right\rangle= n (a^{+}a^{+})^{n} \left|0\right\rangle, 
\end{equation}
so $\hat{N}$ is the particle number operator. Since $[M_{ab},\hat{N} ]=0$, we actually have 
\begin{equation}
	\frac{1}{2}	[O^{-}_{1,0}(y), O^{+}_{1,0}(y)]= \hat{N}
\end{equation}
$\forall\; y \in AdS_{4}$. Moreover, 
\begin{equation}
\frac{1}{2}	[O^{-}_{1,0}(o),[H, O^{+}_{1,0}(o)]]=H,
\end{equation}
\begin{equation}
\frac{1}{2}	[O^{-}_{1,0}(o),[P_{\mu}+K_{\mu}, O^{+}_{1,0}(o)]]=P_{\mu}+K_{\mu},
\end{equation}
or 
\begin{equation}
\frac{1}{2}	[O^{-}_{1,0}(o),[M_{0a}, O^{+}_{1,0}(o)]]=M_{0a}=-\frac{i}{2}	[O^{-}_{1,0}(o),D_{a}O^{+}_{1,0}(o)]. 
\end{equation}
$\forall\; y \in AdS_{4}$, 
\begin{equation}
\frac{1}{2}	[O^{-}_{1,0}(y),[M_{0a}(y), O^{+}_{1,0}(y)]]=M_{0a}(y)=-\frac{i}{2}	[O^{-}_{1,0}(y),D_{a}O^{+}_{1,0}(y)]. 
\end{equation}
$M_{0a}(y)$ and $M_{0a}(o)=M_{0a}$ are related by the $SO(3,2)$ transformation. $D_{a} = e^{u}_{a}(y)\partial_{u}$. Based on the state correspondence, there are local operators $\Phi^{+}_{1,0}(y)$ in $AdS$ theory,
\begin{equation}
\frac{1}{2}	[\Phi^{-}_{1,0}(y),[M_{0a}(y), \Phi^{+}_{1,0}(y)]]=M_{0a}(y)=-\frac{i}{2}	[\Phi^{-}_{1,0}(y),D_{a}\Phi^{+}_{1,0}(y)], 
\end{equation}
and especially, 
\begin{equation}
\frac{1}{2}	[\Phi^{-}_{1,0}(o),[H, \Phi^{+}_{1,0}(o)]]=H = -\frac{i}{2}	[\Phi^{-}_{1,0}(o), \partial_{t}\Phi^{+}_{1,0}(o)]. 
\end{equation}

Finally, since the Hamiltonian and the corresponding spectrum and states are in one-to-one correspondence in AdS and CFT, the partition functions $Z= \textnormal{Tr}e^{-\beta H}$ should also be equal in both sides \cite{ft67y,5t}. We have considered the CFT with the $SO(3,2)$ invariance. One may add the suitable Noether coupling $h^{i_{1}\cdots i_{s}}(x)O_{i_{1}\cdots i_{s}}(x)$ into the Lagrangian so that the $SO(3,2)$ symmetry is broken to $R^{1}$. The CFT will have the modified Hamiltonian $\tilde{H}$, for which, one can also find the spectrum $\{\omega\}$ and states $\{\left|\omega \right\rangle\}$. Then the question is whether there is also a gauge dual for the modified CFT. When $h^{i_{1}\cdots i_{s}}=0$, the dual theory is the higher spin gravity in $AdS_{4}$. It is unlikely that the gauge/gravity correspondence will suddenly cease just because $h^{i_{1}\cdots i_{s}}$ is a little away from $0$. A natural guess is that the dual theory is the higher spin gravity on a new background that is a perturbation of $AdS_{4}$. Let $G(x,\rho)$ be the modified background fields, then for the gauge/gravity correspondence to be valid, $G(x,\rho)$ should be entirely induced from $h^{i_{1}\cdots i_{s}}(x)$, i.e. $G =G(h)$. On CFT side, with $h^{i_{1}\cdots i_{s}}(x)$ the coupling constants, RG flow will then give the entire $G(x,\rho)$. On AdS side, one may take $h^{i_{1}\cdots i_{s}}(x)$ as the boundary condition to get $G(x,\rho)$ by solving the equation of motion. The partition functions on both sides are still expected to be equal, so $Z_{CFT}(h) = Z_{AdS}(G) =Z_{AdS}[G(h)] $. In the generic case, the $R^{1}$ symmetry is also broken, and so, there is no conserved $\tilde{H}$. Nevertheless, the correspondence is still valid with $G =G(h)$ and $Z_{CFT}(h) = Z_{AdS}[G(h)]$.

We have seen that the $SO(3,2)$ transformation of the boundary $AdS$ operators is the same as the conformal transformation of the primary CFT operators with the definite conformal dimension. With the $AdS$ vacuum identified with the CFT vacuum, the boundary $AdS$ operator identified with the corresponding CFT operator, there will be
\begin{equation}\label{sed}
	\left\langle \Omega\right|\Phi_{\lambda,s}(x_{1})\Phi_{\lambda,s}(x_{2})\cdots \Phi_{\lambda,s}(x_{n})\left|\Omega\right\rangle = \left\langle  0 \right|O_{\lambda,s}(x_{1})O_{\lambda,s}(x_{2})\cdots O_{\lambda,s}(x_{n})\left|0\right\rangle,
\end{equation}
where the left and the right hand side of (\ref{sed}) are correlation functions of the boundary operators in $AdS$ and the corresponding operators in CFT respectively. This is in agreement with Witten's prescription. Suppose $J_{\lambda,s}(x)$ are sources at the $AdS$ boundary and in CFT coupling with $\Phi_{\lambda,s}(x)$ and $O_{\lambda,s}(x)$, there will be
\begin{equation}
	W_{AdS}(J_{\lambda,s}) = W_{CFT}(J_{\lambda,s}),
\end{equation}
where 
\begin{equation}\label{208}
	e^{iW_{AdS}(J_{\lambda,s})} = \int D\Phi_{\lambda,s}(X)\;e^{iS_{AdS}[\Phi_{\lambda,s}(X)]+i \int d^{3}x\;\Phi_{\lambda,s}(x) J_{\lambda,s}(x)},
\end{equation}
\begin{equation}
	e^{iW_{CFT}(J_{\lambda,s})} = \int D\phi(x)\;e^{iS_{CFT}[\phi(x)]+i \int d^{3}x\;O_{\lambda,s}(x) J_{\lambda,s}(x)}.
\end{equation}
$X$ is the coordinate in $AdS$, $\Phi_{\lambda,s}(X)$ and $\phi(x)$ represent fields in $AdS$ theory and CFT respectively. In (\ref{208}), $J_{\lambda,s}(x)$ acts as the source at the $AdS$ boundary. Strictly speaking, (\ref{208}) is different from  
\begin{equation}\label{cft1}
	e^{iW_{AdS}(J_{\lambda,s})} = \int D\Phi_{\lambda,s}(X)\;e^{iS_{AdS}[\Phi_{\lambda,s}(X)]|_{\Phi_{\lambda,s}(x) = J_{\lambda,s}(x)}},
\end{equation}
in which, the boundary field is fixed to be $J_{\lambda,s}(x)$. After the Legendre transformation, (\ref{208}) becomes
\begin{equation}
W_{AdS}(J_{\lambda,s}) = \Gamma[\Phi_{\lambda,s}^{J}(X)]+ 	\int d^{3}x\;\Phi_{\lambda,s}^{J}(x) J_{\lambda,s}(x), 
\end{equation}
where
\begin{equation}\label{wedf}
	\Phi_{\lambda,s}^{J}(X)=\frac{\delta}{\delta J_{\lambda,s}(X)}W_{AdS}[J_{\lambda,s}(X)]|_{\bar{J}_{\lambda,s}(X)} 
\end{equation}
is the functional of $J_{\lambda,s}(x)$. $\bar{J}_{\lambda,s}(X)=J_{\lambda,s}(x)\delta (X-x)$. $\Gamma[\Phi_{\lambda,s}^{J}(X)]$ is the effective action of the AdS theory. 
\begin{equation}
	i \Gamma[\Phi_{\lambda,s}^{J}(X)]=\int_{1PI, connected} D\tilde{\Phi}_{\lambda,s}(X)\;e^{iS_{AdS}[\Phi^{J}_{\lambda,s}(X)+\tilde{\Phi}_{\lambda,s}(X)]}. 
\end{equation}
\begin{equation}
	\frac{\delta \Gamma[\Phi_{\lambda,s}]}{\delta \Phi_{\lambda,s}^{\bar{J}}(X)} = -\bar{J}_{\lambda,s}(X). 
\end{equation}
In the bulk, $\delta \Gamma[\Phi_{\lambda,s}]/\delta \Phi_{\lambda,s}^{J}(X)=0$, so $\Phi_{\lambda,s}^{J}(X)$ is the on-shell solution of the effective action generated by the boundary source $J_{\lambda,s}(x)$. From (\ref{wedf}), $\Phi_{\lambda,s}^{J}(X)$ can be solved as
\begin{eqnarray}\label{cft}
\Phi_{\lambda,s}^{J}(X) \nonumber &=&\int d^{3}x_{1}	\Delta(X,x_{1})J_{\lambda,s}(x_{1})+\frac{1}{2!}\int d^{3}x_{1}d^{3}x_{2}	\Delta(X,x_{1},x_{2})J_{\lambda,s}(x_{1})J_{\lambda,s}(x_{2})  \\ &+&  \frac{1}{3!}\int d^{3}x_{1}d^{3}x_{2}d^{3}x_{3}	\Delta(X,x_{1},x_{2},x_{3})J_{\lambda,s}(x_{1})J_{\lambda,s}(x_{2})J_{\lambda,s}(x_{3})+ \cdots,
\end{eqnarray}
where 
\begin{equation}
	\Delta(X,x_{1},\cdots,x_{n}) =\lim_{X_{1}\rightarrow x_{1},\cdots,X_{n}\rightarrow x_{n}} \Delta(X,X_{1},\cdots,X_{n}). 
\end{equation}
\begin{equation}
\Delta(X_{1},\cdots,X_{n})=\frac{\delta^{n}W_{AdS}[J_{\lambda,s}(X)]}{\delta J_{\lambda,s}(X_{1}) \cdots \delta J_{\lambda,s}(X_{n})}	|_{AdS}
\end{equation}
is the connected n-point Green's function of the theory on AdS background. In the simplest situation, if
\begin{equation}
	S_{AdS}[\Phi_{1,0}(X)] = \int dX \; \sqrt{g} (\frac{1}{2}g^{\mu\nu}\partial_{\mu}\Phi_{1,0}\partial_{\nu}\Phi_{1,0}+\frac{1}{2}m^{2}\Phi^{2}_{1,0}), 
\end{equation}
\begin{equation}
	\Gamma[\Phi_{1,0}(X)] =- \int dX \sqrt{g} \Phi_{1,0} \square_{g}  \Phi_{1,0}  , 
\end{equation}
\begin{equation}
	\Phi_{1,0}^{J}(X)  =\int d^{3}x	\; \Delta_{0}(X,x)J_{1,0}(x). 
\end{equation}
\begin{eqnarray}\label{adsx}
W_{AdS}[J_{1,0}(x)] \nonumber &=&\Gamma[\Phi_{1,0}^{J}(X)]+ 	\int d^{3}x\;\Phi_{1,0}^{J}(x) J_{1,0}(x)=		S_{AdS}[\Phi_{1,0}^{J}(X)]  \\ &=& \int d^{3}x_{1}d^{3}x_{2}\;\Delta_{0}(x_{1},x_{2})J_{1,0}(x_{1}) J_{1,0}(x_{2}) , 
\end{eqnarray}
which is in agreement with the standard calculation. $\Delta_{0}(X,x)$ also makes $\Phi_{1,0}^{J}(X) \rightarrow J_{1,0}(x)$ at the boundary. With the interpretation (\ref{cft1}), (\ref{adsx}) is a classical approximation, but with the interpretation (\ref{208}), (\ref{adsx}) is exact.

To conclude, the state correspondence indicates that
\begin{equation}
 W_{CFT}[J_{\lambda,s}(x)]=	\Gamma[\Phi_{\lambda,s}^{J}(X)]+ 	\int d^{3}x\;\Phi_{\lambda,s}^{J}(x) J_{\lambda,s}(x)
\end{equation}
with $\Phi_{\lambda,s}^{J}(X)$ solved via (\ref{cft}). $\Phi_{\lambda,s}^{J}(X)$ is the on-shell solution of the effective action $\Gamma $ generated by the source $J_{\lambda,s}(x)$ at the boundary. When $J_{\lambda,s}(x)=0$, we have
\begin{equation}
W_{CFT}(0)=	W_{AdS}(0) = \Gamma(0). 
\end{equation}
The free energy of CFT is equal to the free energy of the $AdS$ theory as well as the effective action of the $AdS$ background. In this sense, CFT with $J_{\lambda,s}=0$ is quite special. If CFT with the generic $J_{\lambda,s}(x)$ stands on equal footing, for each $J_{\lambda,s}(x)$, we need to find background $H_{J}$ so that  
\begin{equation}
	e^{iW_{AdS}(J_{\lambda,s})} = \int D\Phi_{\lambda,s}(X)e^{iS_{H_{J}}[\Phi_{\lambda,s}(X)]},
\end{equation}
where $S_{H_{J}}$ is the classical action of the bulk theory on background $H_{J}(X)$, which in general cannot be taken as the perturbation of $AdS$. There will be 
\begin{equation}
W_{CFT}[J_{\lambda,s}(x)]=	W_{H_{J}}(0) = \Gamma_{H_{J}}(0). 
\end{equation}
Just as the $AdS$ case, for bulk theory on $H_{J}(X)$, the boundary correlation function can be identified with the correlation function of the CFT on $3d$ background $J_{\lambda,s}(x)$.

Now we have two kinds of the bulk fields $\Phi_{\lambda,s}^{J}(X)$ and $H_{J}(X)$, both are determined by $J_{\lambda,s}(x)$. The two are closely related. Especially, for the boundary 1-point function,
\begin{equation}
 \Phi^{J}_{\lambda,s}(x)= \int D\tilde{\Phi}_{\lambda,s}(X) \tilde{\Phi}_{\lambda,s}(x)e^{iS_{AdS}[-\Phi^{J}_{\lambda,s}(X)+ \tilde{\Phi}_{\lambda,s}(X)]}=\frac{\delta}{\delta J_{\lambda,s}(x)}W_{CFT}(J_{\lambda,s})|_{J_{\lambda,s}(x)}. 
\end{equation}
However, generically, $S_{H_{J}}[\tilde{\Phi}_{\lambda,s}(X)] \neq S_{AdS}[-\Phi^{J}_{\lambda,s}(X)+ \tilde{\Phi}_{\lambda,s}(X)]$, since the former cannot be obtained by the simple perturbation. It is also desirable to get the bulk field from CFT with $J_{\lambda,s}(x)$. One recipe is the RG flow. It is also hoped that there is a way to rewrite the partition function of the CFT so that $W_{CFT}[J_{\lambda,s}(x)]=\mathcal{W}_{CFT}[f_{J}(X)]$, where $f_{J}(X)$ is a $5d$ function induced from $J_{\lambda,s}(x)$, which may be $H_{J}(X)$ or $\Phi^{J}_{\lambda,s}(X)$.

\section{Conclusion}

We have discussed the radial quantization of the $3d$ free $O(N)$ vector model, which is supposed to be dual to the higher spin gravity in $AdS_{4}$. Higher spin charges get the oscillator realization in CFT. We only calculated two special classes of charges $P^{s}_{\mu_{1} \cdots \mu_{n}}$ and $K^{s}_{\mu_{1} \cdots \mu_{n}}$ explicitly, but it is straightforward to get the whole set of conserved charges composing the higher spin algebra. The discussion can be directly extended to the $D$ dimensional CFT.

The non-pertubative quantum higher spin gravity in $AdS_{4}$ should have a set of the basis composed by the physical single particle states $\left|n,l;\lambda,s\right\rangle$ as well as the multi-particle states. In CFT, the fock states of the higher spin gravity are realized as 
\begin{equation}
	\left|n,l;s+1,s\right\rangle = O^{n}_{\mu_{1}\cdots \mu_{l};\;i_{1}\cdots i_{s}}(0) \left|0\right\rangle
\end{equation}
and
\begin{equation}
	\left|n_{1},l_{1};s_{1}+1,s_{1}:\cdots:n_{m},l_{m};s_{m}+1,s_{m}\right\rangle = O^{n_{1}}_{\mu^{1}_{1}\cdots \mu^{1}_{l_{1}};\;i^{1}_{1}\cdots i^{1}_{s_{1}}}(0) \cdots O^{n_{m}}_{\mu^{m}_{1}\cdots \mu^{m}_{l_{m}};\;i^{m}_{1}\cdots i^{m}_{s_{m}}}(0) \left|0\right\rangle.
\end{equation}
It is easy to prove that the 1-particle Hilbert space of the higher spin gravity in AdS forms the irreducible representation of the higher spin algebra. The conclusion holds for the $D$ dimensional CFT and $AdS_{D+1}$. The CFT realization of the fock states automatically carries the right degrees of freedom with no redundancy to remove. As a result, there is no local higher spin symmetry left. We can compute the inner products between the fock states, which are supposed to carry the whole dynamical information of the theory and are expected to be reproduced once the higher spin gravity is quantized.

The vector model/higher spin gravity duality gives the simplest example of the  CFT definition of the quantum gravity. For the generic AdS/CFT correspondence, the difficulty lies in the explicit CFT construction of $O^{n}_{\mu_{1}\cdots \mu_{l};\;i_{1}\cdots i_{s}}(0)$ but the spirit remains the same. Except for the spectrums and the corresponding states, one may also want to get the CFT realization of $\Psi(x,\rho)$, the Heisenberg operators for the fundamental fields of the AdS theory. This is possible as long as one bulk $\Psi(o)$ gets the CFT realization. However, it is not obvious which operator in CFT should be identified as $\Psi(o)$. The operators at hand are the primary operators with the definite conformal dimensions that could be taken as the boundary limit of some AdS operators. With the boundary values given, the bulk extension is not unique. One can always find a consistent bulk extension by solving the free field equation, but $\Psi(x,\rho)$ satisfies the free field equation only in free theory limit. Except for $\Psi(x,\rho)$, the other set of operators of interest is $\Phi^{+}(x,\rho)$, the creation operators for physical particles in AdS. $\Phi^{+}(x,\rho)$ satisfies the free field equation and could be explicitly constructed in CFT. In a non-perturbative treatment of the quantum higher spin gravity in AdS, we will still try to find $\Phi^{+}(x,\rho)$ at the end, because it is $\Phi^{+}(x,\rho)$ that is directly related to the observation.

\bigskip
\bigskip

\section*{Acknowledgments}


The work is supported by the Natural Science Foundation of China under grant numbers 10821504, 11075194, 11135003, and 11275246.

\appendix

\section{The calculation of $D$ and $P^{\mu}$ }

\begin{equation}
	D = i \int_{S^{2}} d\Omega \;n_{\mu}n_{\nu} 	T^{\mu\nu}(1, \theta,\varphi), 
\end{equation}
\begin{equation}
n_{\mu}n_{\nu}	T^{\mu\nu}=\frac{3}{4}n_{\mu}\partial^{\mu}\phi n_{\nu}\partial^{\nu}\phi-\frac{1}{4}\partial^{\alpha}\phi\partial_{\alpha}\phi - \frac{1}{8}(\phi n_{\mu}n_{\nu}\partial^{\mu}\partial^{\nu}\phi+ n_{\mu}n_{\nu}\partial^{\mu}\partial^{\nu}\phi  \phi). 
\end{equation}
The direct calculation gives 
\begin{eqnarray}\label{1q}
\nonumber && \int_{S^{2}} d\Omega \; n^{\mu}\partial_{\mu}\phi (1,\theta,\varphi) n^{\nu}\partial_{\nu}\phi (1,\theta,\varphi) \\\nonumber &=& \sum \;\frac{N_{j_{1} \cdots j_{l};\;i_{1} \cdots i_{l}} }{2l+1}  
[l^{2}a^{+}_{j_{1}\cdots i_{l}} a^{+}_{i_{1}\cdots i_{l}}- l(l+1)a_{j_{1}\cdots i_{l}}a^{+}_{i_{1}\cdots i_{l}}
\\&& -l(l+1)a^{+}_{j_{1}\cdots i_{l}} a_{i_{1}\cdots i_{l}}+ (l+1)^{2}a_{j_{1}\cdots i_{l}}a_{i_{1}\cdots i_{l}}]
,
\end{eqnarray}
\begin{equation}\label{2q}
\int_{S^{2}} d\Omega \; \partial^{\mu}\phi (1,\theta,\varphi) \partial_{\mu}\phi (1,\theta,\varphi) = \sum \; N_{j_{1} \cdots j_{l};\;i_{1} \cdots i_{l}} [(l+1)a_{j_{1} \cdots j_{l}} a_{i_{1} \cdots i_{l}}+l a^{+}_{j_{1} \cdots j_{l}} a^{+}_{i_{1} \cdots i_{l}} ] ,	
\end{equation}
\begin{eqnarray}\label{3q}
\nonumber	&& \int_{S^{2}} d\Omega \; \frac{1}{2}[\phi (1,\theta,\varphi)n^{\mu}n^{\nu}\partial_{\nu}\partial_{\mu}\phi (1,\theta,\varphi)+n^{\mu}n^{\nu}\partial_{\nu}\partial_{\mu}\phi (1,\theta,\varphi) \phi (1,\theta,\varphi)]\\\nonumber	 &=&     \sum  \frac{N_{j_{1} \cdots j_{l};\;i_{1} \cdots i_{l}}}{2l+1}[(l+1)(l+2) a_{j_{1}\cdots j_{l}}a_{i_{1}\cdots i_{l}}+(l^{2}+l+1)a^{+}_{j_{1}\cdots j_{l}}a_{i_{1}\cdots i_{l}}\\&+& (l^{2}+l+1)a_{j_{1}\cdots j_{l}}a^{+}_{i_{1}\cdots i_{l}}+l(l-1)a^{+}_{j_{1}\cdots j_{l}}a^{+}_{i_{1}\cdots i_{l}}],
\end{eqnarray}
where we have used 
\begin{equation}
	\int d^{2}\Omega \; Y_{j_{1} \cdots j_{l}}Y_{i_{1} \cdots i_{l'}} =\frac{4\pi\delta_{l,l'}(-1)^{l} (l!)^{2}}{(2l-1)!!(2l+1)!!}\partial_{j_{1}} \cdots \partial_{j_{l}}(r^{2l+1}\partial_{i_{1}} \cdots \partial_{i_{l}}\frac{1}{r})=N_{j_{1} \cdots j_{l};\;i_{1} \cdots i_{l}},
\end{equation}
\begin{equation}
	N_{\mu j_{1} \cdots j_{l};\;\mu i_{1} \cdots i_{l}}=\frac{l+1}{2l+1}N_{j_{1} \cdots j_{l};\;i_{1} \cdots i_{l}},
\end{equation}
\begin{equation}
n^{\mu}	Y_{\mu i_{1}\cdots i_{l}}=\frac{l+1}{2l+1}Y_{i_{1}\cdots i_{l}}. 
\end{equation}

With (\ref{1q})-(\ref{3q}) plugged in, we get 
\begin{eqnarray}
D \nonumber &=& -i \sum \;  \frac{N_{j_{1} \cdots j_{l};\;i_{1} \cdots i_{l}}}{4} [ (2l+1)(a^{+}_{j_{1}\cdots j_{l}}a_{i_{1}\cdots i_{l}}+  a_{j_{1}\cdots j_{l}}a^{+}_{i_{1}\cdots i_{l}})] \\\nonumber &=& - i \sum_{l,m} \;  \frac{1}{4} [ (2l+1)(a^{+}_{l,m}a_{l,m}+  a_{l,m}a^{+}_{l,m})] \\\nonumber &=& - i \sum_{l,m} \;   (l+\frac{1}{2}) a^{+}_{l,m}a_{l,m} + \textnormal{const} \\ &=& - i \sum \;   (l+\frac{1}{2}) N_{j_{1} \cdots j_{l};\;i_{1} \cdots i_{l}} a^{+}_{j_{1}\cdots j_{l}}a_{i_{1}\cdots i_{l}} + \textnormal{const}. 
\end{eqnarray}

\begin{equation}
	P^{\mu}=i \int_{S^{2}} d\Omega \;n_{\nu} 	T^{\mu\nu}(1, \theta,\varphi),
\end{equation}
\begin{equation}
n_{\mu}	T^{\mu\nu}= \frac{3}{8}(n_{\mu}\partial^{\mu}\phi\partial^{\nu}\phi +\partial^{\nu}\phi n_{\mu}\partial^{\mu}\phi)-\frac{1}{4}n^{\nu}\partial^{\alpha}\phi\partial_{\alpha}\phi  - \frac{1}{8}(\phi n_{\mu}\partial^{\mu}\partial^{\nu}\phi +  n_{\mu}\partial^{\mu}\partial^{\nu}\phi \phi). 
\end{equation}
\begin{eqnarray}
\nonumber	&& \int_{S^{2}} d\Omega \;n^{\mu}\partial_{\mu}\phi  \partial_{\nu}\phi +\partial_{\nu}\phi n^{\mu}\partial_{\mu}\phi \\\nonumber &=&  \sum \; (\frac{\sqrt{2l+1}}{\sqrt{2l'+1}})\{-[a^{+}_{j_{1}\cdots j_{l'}} l'- a_{j_{1}\cdots j_{l'}}(l'+1) ]( a_{i_{1}i_{2}\cdots i_{l}} + a^{+}_{i_{1}i_{2}\cdots i_{l}}) Y_{ j_{1}\cdots j_{l'}} Y_{\nu i_{1}\cdots i_{l}} \\ \nonumber &&+ \;  [a^{+}_{j_{1}\cdots j_{l'}} l'- a_{j_{1}\cdots j_{l'}}(l'+1) ]a^{+}_{i_{1}\cdots i_{l}} n_{\nu}Y_{ j_{1}\cdots j_{l'}}  Y_{i_{1}\cdots i_{l}}    \} \\ && + \;
\sum \; (\frac{\sqrt{2l'+1}}{\sqrt{2l+1}})\{-( a_{j_{1}\cdots j_{l'}} + a^{+}_{j_{1}\cdots j_{l'}})[a^{+}_{i_{1}\cdots i_{l}} l- a_{i_{1}\cdots i_{l}}(l+1) ]Y_{\nu j_{1}\cdots j_{l'}} Y_{ i_{1}\cdots i_{l}}  \\ \nonumber &&  +\; a^{+}_{j_{1}\cdots j_{l'}} [a^{+}_{i_{1}\cdots i_{l}} l- a_{i_{1}\cdots i_{l}}(l+1) ]n_{\nu}Y_{ j_{1}\cdots j_{l'}}  Y_{i_{1}\cdots i_{l}}    \}.	 
\end{eqnarray}
\begin{eqnarray}
\nonumber	&& \int_{S^{2}} d\Omega \;n_{\nu}\partial^{\mu}\phi (1,\theta,\varphi)\partial_{\mu}\phi (1,\theta,\varphi) \\\nonumber&=& \sum \sum \; \sqrt{(2l+1)(2l'+1)}[( a_{j_{1}\cdots j_{l'}} + a^{+}_{j_{1}\cdots j_{l'}})( a_{i_{1}\cdots i_{l}} + a^{+}_{i_{1}\cdots i_{l}}) ]  Y_{\mu j_{1}\cdots j_{l'}}Y_{\mu i_{1}\cdots i_{l}}n_{\nu} \\\nonumber&& - \{ a^{+}_{j_{1}\cdots j_{l'}}( a_{i_{1}\cdots i_{l}} + a^{+}_{i_{1}\cdots i_{l}}) \frac{(l+1)\sqrt{2l'+1}}{\sqrt{2l+1}} 
\\\nonumber&& 
+( a_{j_{1}\cdots j_{l'}} + a^{+}_{j_{1}\cdots j_{l'}})a^{+}_{i_{1}\cdots i_{l}} \frac{(l'+1)\sqrt{2l+1}}{\sqrt{2l'+1}}  \\ && - \sqrt{(2l+1)(2l'+1)} a^{+}_{j_{1}\cdots j_{l'}}a^{+}_{i_{1}\cdots i_{l}}\}\; n_{\nu}  Y_{j_{1}\cdots j_{l'}} Y_{i_{1}\cdots i_{l}}.	 
\end{eqnarray}
\begin{eqnarray}
\nonumber	&& \int_{S^{2}} d\Omega \;\phi (1,\theta,\varphi) n^{\mu} \partial_{\nu}\partial_{\mu}\phi (1,\theta,\varphi)+n^{\mu} \partial_{\nu}\partial_{\mu}\phi (1,\theta,\varphi) \phi (1,\theta,\varphi)\\\nonumber  &=& 
\sum\sum \; \sqrt{\frac{2l+1}{2l'+1}} \{( a_{j_{1}\cdots j_{l'}}+  a^{+}_{j_{1}\cdots j_{l'}}  )[(l+2)a_{i_{1}\cdots i_{l}}-(l-1)a^{+}_{i_{1}\cdots i_{l}}]Y_{j_{1}\cdots j_{l'}}Y_{\nu i_{1}\cdots i_{l}}\} \\\nonumber&& +\sum\sum \;(l-1) \sqrt{\frac{2l+1}{2l'+1}} [( a_{j_{1}\cdots j_{l'}}+  a^{+}_{j_{1}\cdots j_{l'}}  )a^{+}_{i_{1}\cdots i_{l}}n_{\nu} Y_{j_{1}\cdots j_{l'}} Y_{i_{1}\cdots i_{l}}   ]  \\\nonumber&& + \sum\sum \; \sqrt{\frac{2l'+1}{2l+1}} \{[(l'+2)a_{j_{1}\cdots j_{l'}}-(l'-1)a^{+}_{j_{1}\cdots j_{l'}}]( a_{i_{1}\cdots i_{l}}+  a^{+}_{i_{1}\cdots i_{l}}  )Y_{\nu j_{1}\cdots j_{l'}}Y_{i_{1}\cdots i_{l}}\} \\&& +\sum\sum \;(l'-1) \sqrt{\frac{2l'+1}{2l+1}}[a^{+}_{j_{1}\cdots j_{l'}} ( a_{i_{1}\cdots i_{l}}+  a^{+}_{i_{1}\cdots i_{l}}  )n_{\nu} Y_{j_{1}\cdots j_{l'}}Y_{i_{1}\cdots i_{l}}   ].	 
\end{eqnarray}
Note that  
\begin{equation}
\int_{S^{2}} d\Omega \; n_{\nu}  Y_{j_{1}\cdots j_{l'}} Y_{i_{1}\cdots i_{l}}=0, 	
\end{equation}
\begin{equation}
	\int_{S^{2}} d\Omega \; n_{\nu}  Y_{\mu j_{1}\cdots j_{l'}} Y_{\mu i_{1}\cdots i_{l}} =0,
\end{equation}
unless $l-l'=\pm 1$. Without loss of generality, we can let $l'=l+1$, $P_{\nu}$ becomes 
\begin{eqnarray}\label{pl1}
\nonumber P_{\nu}  &=&- \frac{i}{4}\int_{S^{2}} d\Omega \; \sum \; \sqrt{(2l+1)(2l+3)}\; n_{\nu}  Y_{j_{1}\cdots j_{l+1}} Y_{i_{1}\cdots i_{l}}\; [-\frac{4(l+2)}{2l+3}a^{+}_{j_{1}\cdots j_{l+1}}a^{+}_{i_{1}\cdots i_{l}}\\\nonumber && \;\;\;\;\;\;\;\;\;\;\;\;\;\;\;\;\;\;\;\;\;\;\;\;\;\;\;\;\;\;\;\;\;\;\;\;\;\;\;\;\;\;\;\;\;\;\;\;\;+ \;  \frac{2l+1}{2l+3} a_{j_{1}\cdots j_{l+1}}a^{+}_{i_{1}\cdots i_{l}} +  a^{+}_{j_{1}\cdots j_{l+1}}a_{i_{1}\cdots i_{l}} ]\\\nonumber && + \; \sqrt{\frac{2l+1}{2l+3}}Y_{j_{1}\cdots j_{l+1}} Y_{\nu i_{1}\cdots i_{l}}\;[2(l+2)(a^{+}_{j_{1}\cdots j_{l+1}}a^{+}_{i_{1}\cdots i_{l}}-a_{j_{1}\cdots j_{l+1}}a_{i_{1}\cdots i_{l}})\\\nonumber &&\;\;\;\;\;\;\;\;\;\;\;\;\;\;\;\;\;\;\;\;\;\;\;\;\;\;\;\;\;\;\;\;\;\;\;\;\;\; + \; (4l+5)(a^{+}_{j_{1}\cdots j_{l+1}}a_{i_{1}\cdots i_{l}}- a_{j_{1}\cdots j_{l+1}} a^{+}_{i_{1}\cdots i_{l}})]
 \\ \nonumber &&+ \; 2 \sqrt{(2l+1)(2l+3)}n_{\nu}  Y_{\mu j_{1}\cdots j_{l+1}} Y_{\mu i_{1}\cdots i_{l}}(a_{j_{1}\cdots j_{l+1}}a_{i_{1}\cdots i_{l}}+a_{j_{1}\cdots j_{l+1}}a^{+}_{i_{1}\cdots i_{l}}\\\nonumber && \;\;\;\;\;\;\;\;\;\;\;\;\;\;\;\;\;\;\;\;\;\;\;\;\;\;\;\;\;\;\;\;\;\;\;\;\;\;\;\;\;\;\;\;\;\;\;\;\;\;\;\;\;\; + \; a^{+}_{j_{1}\cdots j_{l+1}}a_{i_{1}\cdots i_{l}}+a^{+}_{j_{1}\cdots j_{l+1}}a^{+}_{i_{1}\cdots i_{l}})
.	 
\end{eqnarray}
Moreover, since 
\begin{equation}
\int_{S^{2}} d\Omega \; n_{\nu}  Y_{j_{1}\cdots j_{l+1}} Y_{i_{1}\cdots i_{l}}=\int_{S^{2}} d\Omega \; Y_{j_{1}\cdots j_{l+1}} Y_{\nu i_{1}\cdots i_{l}}, 	
\end{equation}
\begin{equation}
	\int_{S^{2}} d\Omega \; n_{\nu}  Y_{\mu j_{1}\cdots j_{l+1}} Y_{\mu i_{1}\cdots i_{l}} =\frac{l+2}{2l+3} \int_{S^{2}} d\Omega \; Y_{j_{1}\cdots j_{l+1}} Y_{\nu i_{1}\cdots i_{l}},
\end{equation}
(\ref{pl1}) can be further simplified into
\begin{equation}
P_{\nu}=- i 	\sum \; \sqrt{(2l+1)(2l+3)} N_{j_{1}\cdots j_{l+1};\;\nu i_{1}\cdots i_{l}}a^{+}_{j_{1}\cdots j_{l+1}}a_{i_{1}\cdots i_{l}}. 
\end{equation}

\bibliographystyle{plain}

\end{document}